\documentclass[journal]{IEEEtran}

\usepackage{cite}

\ifCLASSINFOpdf
   \usepackage[pdftex]{graphicx}
   \graphicspath{{../pdf/}{../jpeg/}}
   \DeclareGraphicsExtensions{.pdf,.jpeg,.png}
\else
   \usepackage[dvips]{graphicx}
   \graphicspath{{../eps/}}
   \DeclareGraphicsExtensions{.eps}
\fi

\usepackage{amsmath,bm}

\usepackage{algorithmic}

\usepackage{array}

\ifCLASSOPTIONcompsoc
  \usepackage[caption=false,font=normalsize,labelfont=sf,textfont=sf]{subfig}
\else
  \usepackage[caption=false,font=footnotesize]{subfig}
\fi

\usepackage{fixltx2e}

\usepackage{url}

\usepackage{soul}
\usepackage{color}
\usepackage{setspace}
\usepackage{multirow}
\usepackage[flushleft]{threeparttable}

\usepackage{bm}
\usepackage{amsbsy}
\usepackage{float}
\usepackage{mathtools}

\makeatletter
\def\endthebibliography{%
  \def\@noitemerr{\@latex@warning{Empty `thebibliography' environment}}%
  \endlist
}
\makeatother

\begin{document}

\setstcolor{red}

\newcommand{\etal}{\textit{et al}. }
\newcommand{\ie}{\textit{i}.\textit{e}. }
\newcommand{\eg}{\textit{e}.\textit{g}. }

\newcommand{\matJ}[0]{\widetilde{\mathbf{J}}}
\newcommand{\matM}[0]{\mathbf{M}}
\newcommand{\matX}[0]{\mathbf{X}}
\newcommand{\matE}[0]{\mathbf{E}}
\newcommand{\matH}[0]{\widetilde{\mathbf{H}}}

\newcommand{\opLE}[0]{\mathcal{L}_E}
\newcommand{\opLH}[0]{\mathcal{L}_H}
\newcommand{\opK}[0]{\mathcal{K}}
\newcommand{\opKpv}[0]{\widetilde{\mathcal{K}}}

\newcommand{\matf}[0]{\mathbf{f}}
\newcommand{\matPhi}[0]{\mathbf{\Phi}}

\newcommand{\matL}[0]{\mathbf{L}}
\newcommand{\matK}[0]{\mathbf{K}}

\newcommand{\nx}[0]{\hat{n} \times}
\newcommand{\nxnx}[0]{\hat{n} \times \hat{n} \times}

\newcommand{\vect}[1]{\boldsymbol{#1}}
\newcommand{\matr}[1]{\mathbf{#1}}
\newcommand{\vr}[0]{\vect{r}}

\newcommand{\fE}[0]{\matr{E}}
\newcommand{\fH}[0]{\matr{H}}

\newcommand{\IntOmega}[0]{\int_{\omega_0 - \Delta \omega}^{\omega_0 + \Delta \omega}}

\title{Wigner-Smith Time Delay Matrix for Electromagnetics: Systems with \\Material Dispersion and Losses}

\author{Yiqian~Mao,
	  Utkarsh~R.~Patel
        and~Eric~Michielssen*,~\IEEEmembership{Fellow,~IEEE}
\thanks{Y. Mao, U. R. Patel and E. Michielssen are with the Department of Electrical Engineering and Computer Science, University of Michigan, Ann Arbor, MI 48109 USA. (e-mail: yqmao@umich.edu; utkarsh.patel@alum.utoronto.ca; emichiel@umich.edu).}
}


\maketitle

\begin{abstract}
The Wigner-Smith (WS) time delay matrix relates a system's scattering matrix to its frequency derivative and gives rise to so-called WS modes that experience well-defined group delays when interacting with the system. For systems composed of nondispersive and lossless materials, the WS time delay matrix previously was shown to consist of volume integrals of energy-like densities plus correction terms that account for the guiding, scattering, or radiating characteristics of the system. This study extends the use of the WS time delay matrix to systems composed of dispersive and lossy materials. Specifically, it shows that such systems' WS time delay matrix can be expressed by augmenting the previously derived expressions with terms that account for the dispersive and lossy nature of the system, followed by a transformation that disentangles effects of losses from time delays. Analytical and numerical examples demonstrate the new formulation once again allows for the construction of frequency stable WS modes that experience well-defined group delays upon interacting with a system.
\end{abstract}

\begin{IEEEkeywords}
Wigner-Smith time delay, group delay, dispersive medium, lossy medium. 
\end{IEEEkeywords}

\IEEEpeerreviewmaketitle

\section{Introduction}
\label{sec:intro}

\IEEEPARstart{W}{igner-Smith} (WS) time delay concepts \cite{Smith1960Lifetime} have been adopted in many branches of wave physics, including quantum physics, electromagnetics, optics, and acoustics. Applications include the characterization of principle modes in fibre optics \cite{Carpen2015Obs}, the study of molecular ionization \cite{Hockett2016time}, wavefront shaping in disordered media \cite{Ambichl2017Focus}, and the construction of particle-like wave packets with minimal dispersion \cite{Gerard2016part}; for other recent applications, see \cite{Gallmann_2017, Brandstotter_2019, Gerardin2014full, Bohm_2018, Durand_2019, Texier_2016, Hougne2021demand, Chen2021general}. Different definitions of the WS time delay matrix for a system with scattering matrix $\matr{S}$ have appeared in the literature, including \cite{Hougne2021demand}
\begin{align}
\label{eq:Q_def1}
\matr{Q}_V = j \matr{S}^\dag \frac{\partial \matr{S}}{\partial \omega}
\end{align}
and \cite{Chen2021general}
\begin{align}
\label{eq:Q_def2}
\matr{Q} = j \matr{S}^{-1} \frac{\partial \matr{S}}{\partial \omega} \,.
\end{align}

For lossless and reciprocal electromagnetic systems, the scattering matrix $\matr{S}$ is unitary, \ie $\matr{S}^\dag = \matr{S}^{-1}$, implying  $\matr{Q} = \matr{Q}_V$. Recently, Patel and Michielssen \cite{Patel2020WS1,Patel2020WS2} showed that for such systems $\matr{Q}$ may be expressed in terms of volume integrals of energy-like densities plus correction terms that depend on the guiding, scattering, or radiating nature of the system. They furthermore developed efficient techniques for computing $\matr{Q}$ using integral equation methods, reducing the dimensionality of the integrals to be evaluated from three to two. Finally, they illustrated that  so-called WS modes obtained by diagonalizing $\matr{Q}$ oftentimes untangle wave and scattering phenomena experiencing well-defined group delays corresponding to $\matr{Q}$'s eigenvalues. Unfortunately, the restriction of the methods in \cite{Patel2020WS1,Patel2020WS2} to systems composed of nondispersive and lossless materials limits their use in many real-world applications. Few studies to date have extended the use of WS methods to such systems.  An exception is recent work by Chen \textit{et al.} \cite{Chen2021general} that analyzes WS time delays in lossy systems by utilizing the pole expansion of $\matr{S}$ and analyzing the average of $\matr{Q}$'s eigenvalues through Eq.~\eqref{eq:Q_def2}.

This study extends \cite{Patel2020WS1} to systems with material dispersion and losses. Its specific contributions are threefold:
\begin{enumerate}
\item It shows that $\matr{Q}_V$ defined via Eqs.~\eqref{eq:Q_def1} can be expressed in terms of the volume integrals of energy-like quantities in \cite{Patel2020WS1} plus additional correction terms that account for the dispersive and lossy nature of the system. Unfortunately, the eigendecomposition of $\matr{Q}_V$ produces limited insights into the system; specifically, its eigenvalues incorporate effects of both time delays and losses.
\item It shows that the eigenvectors and the real parts of eigenvalues of $\matr{Q}= j \matr{S}^{-1} \matr{S}' = (\matr{S}^\dag \matr{S})^{-1} \matr{Q}_{V}$ also diagonalize $\matr{S}$, and hence retain the physical interpretation of WS modes and time delays, respectively. The simultaneous diagonalization of $\matr{Q}$ and $\matr{S}$ implies that the WS modes are fully decoupled and frequency stable. 
\item It uses analytical and numerical examples to illustrate the effects of material dispersion and losses on the eigenstates of $\matr{Q}$, and further highlights the difference between $\matr{Q}$ and $\matr{Q}_V$.
\end{enumerate}

Contributions 1, 2, and 3 are detailed in Sections II, III, and IV below.  Throughout this paper, $f$ denotes frequency and $\omega = 2 \pi f$ denotes angular frequency.  In addition, $'$ denotes $\partial/\partial \omega$, and $^*$, $^T$, and $^\dag$ denote conjugate, transpose, and conjugate transpose operations.

\section{WS Relationship}
\label{sec:em}

\subsection{Setup}

\begin{figure}[H]
\centering\includegraphics[width=8.0cm]{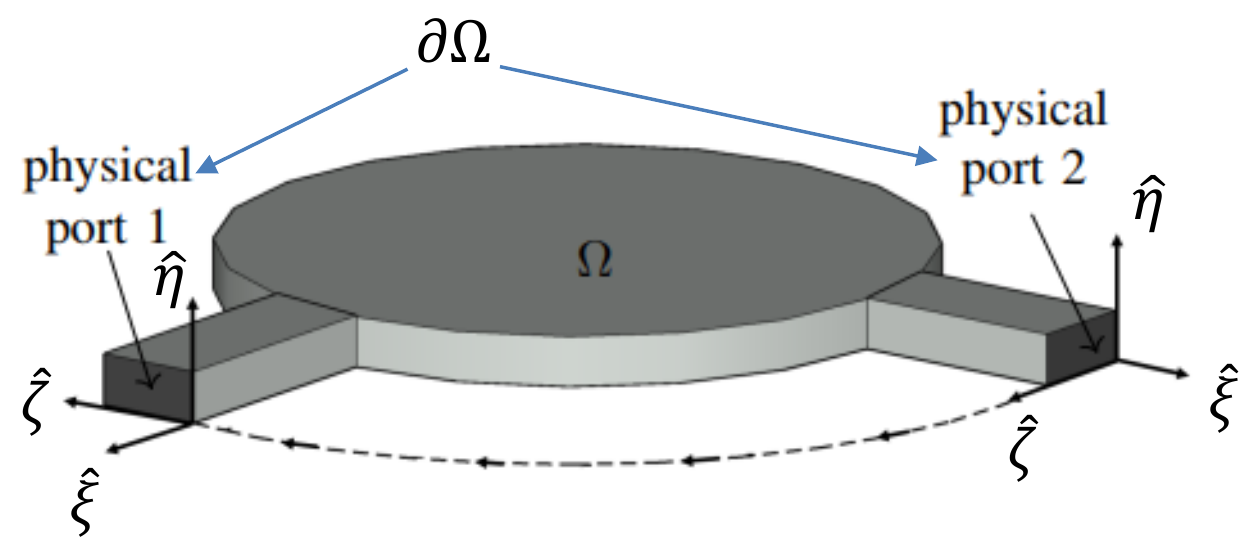}
\caption{Guiding system as in \cite{Patel2020WS1} but filled with dispersive and lossy material.}
\label{fig:SetupGuiding}
\end{figure}

Let $\Omega$ and $\partial \Omega$ denote the volume and port surfaces of a guiding system composed of cavities and waveguides with perfect electrically conducting (PEC) walls (Fig.~\ref{fig:SetupGuiding}). Let $\varepsilon(\vr,\omega)$ and $\mu(\vr,\omega)$ denote the permittivity and permeability of the dispersive and lossy medium that fills $\Omega$ and let $(\eta,\zeta)$ denote a Cartesian parameterization of $\partial \Omega$.  It is assumed that $\partial \Omega$ is removed from all waveguide material and geometric discontinuities, and that  $\varepsilon(\vr,\omega)$ and $\mu(\vr,\omega)$ are homogeneous, lossless, and nondispersive near $\partial \Omega$.  With these assumptions, waveguide fields near $\partial \Omega$ can be expanded in a set of $M$ propagating modes with real, frequency independent, and orthonormal transverse profiles $\bm{\mathcal{X}}_p(\eta,\zeta)$, propagation constants $\beta_p(\omega)$, and impedances $Z_p(\omega)$, $p=1,…,M$ (See Appendix A in \cite{Patel2020WS1}).

Assume the system is excited by the $p$-th incoming mode with transverse field components near $\partial \Omega$ given by
\begin{subequations}
\label{eq:incoming}
\begin{align}
\fE_{p,\|}^i (\vr,\omega) &= \sqrt{Z_p} e^{j \beta_p(\omega) \xi} \bm{\mathcal{X}}_p(\eta,\zeta) \\
\fH_{p,\|}^i (\vr,\omega) &= \frac{1}{\sqrt{Z_p}} e^{j \beta_p(\omega) \xi} (- \hat{\xi} \times \bm{\mathcal{X}}_p(\eta,\zeta)) \,
\end{align}
\end{subequations}
where $\xi$ is the distance away from $\partial \Omega$ outside $\Omega$. Going forward, spatial and frequency dependencies will be omitted whenever possible to simplify notation. 
The outgoing transverse fields near $\partial \Omega$ are 
\begin{subequations}
\begin{align}
\fE_{p,\|}^o (\vr,\omega) &= \sum_{m=1}^{M} S_{mp} \sqrt{Z_m} e^{- j \beta_m \xi} \bm{\mathcal{X}}_m \\
\fH_{p,\|}^o (\vr,\omega) &= - \sum_{m=1}^{M} S_{mp} \frac{1}{\sqrt{Z_m}} e^{- j \beta_m \xi} (- \hat{\xi} \times \bm{\mathcal{X}}_m)
\end{align}
\end{subequations}
where $S_{mp}$ is the system's scattering coefficient characterizing coupling from mode $p$ to mode $m$. For $\vr$ near $\partial \Omega$, the total transverse fields are
\begin{subequations}
\begin{align}
\fE_{p,\|} &= \fE_{p,\|}^i + \fE_{p,\|}^o \label{eq:tot_E_transv} \\
\fH_{p,\|} &= \fH_{p,\|}^i + \fH_{p,\|}^o \,. \label{eq:tot_H_transv}
\end{align}
\end{subequations}

\subsection{Expressions for $\matr{Q}_V$ and $\matr{Q}$}

Let $\fE_{p,q}(\vr,\omega)$ and $\fH_{p,q}(\vr,\omega)$ denote total fields in $\Omega$ excited by the $p/q$-th incoming mode. The frequency derivatives of Maxwell's equations for fields $\fE_p(\vr,\omega)$ and $\fH_p(\vr,\omega)$ are
\begin{subequations}
\begin{align}
    \nabla \times \fH_p' &= j (1+\omega\frac{\varepsilon'}{\varepsilon}) \varepsilon \fE_p + j \omega \varepsilon \fE_p' \label{eq:Helmholtz_dw_E} \\
    \nabla \times \fE_p' &= - j (1+\omega\frac{\mu'}{\mu}) \mu \fH_p - j \omega \mu \fH_p'\,. \label{eq:Helmholtz_dw_H}
\end{align}
\end{subequations}
Likewise, the complex conjugates of Maxwell's equations for fields $\fE_q(\vr,\omega)$ and $\fH_q(\vr,\omega)$ are
\begin{subequations}
\begin{align}
    \nabla \times \fH_q^* &= - j \omega \varepsilon^* \fE_q^* \label{eq:Helmholtz_conj_E} \\
    \nabla \times \fE_q^* &= j \omega \mu \fH_q^*\,. \label{eq:Helmholtz_conj_H} 
\end{align}
\end{subequations}
Adding the dot-product of $\frac{1}{2} \frac{\varepsilon}{\varepsilon^*} \fE_p'$ and Eq.~\eqref{eq:Helmholtz_conj_E} to the dot-product of $\frac{1}{2} \fE_q^*$ and Eq.~\eqref{eq:Helmholtz_dw_E} yields
\begin{align}
\frac{1}{2} \frac{\varepsilon}{\varepsilon^*} \fE_p' \cdot \nabla \times \fH_q^* + \frac{1}{2}\fE_q^* \cdot \nabla \times \fH_p' = \frac{j}{2} (1+\omega\frac{\varepsilon'}{\varepsilon}) \varepsilon \fE_q^* \cdot \fE_p \,. \label{eq:Helmholtz_add1}
\end{align}
Similarly, adding the dot-product of $\frac{1}{2} \frac{\mu}{\mu^*} \fH_p'$ and Eq.~\eqref{eq:Helmholtz_conj_H} to the dot-product of $\frac{1}{2} \fH_q^*$ and Eq.~\eqref{eq:Helmholtz_dw_H} yields
\begin{align}
\frac{1}{2} \frac{\mu}{\mu^*} \fH_p' \cdot \nabla \times \fE_q^* + \frac{1}{2}\fH_q^* \cdot \nabla \times \fE_p' \nonumber \\
= - \frac{j}{2} (1+\omega\frac{\mu'}{\mu}) \mu \fH_q^* \cdot \fH_p \,. \label{eq:Helmholtz_add2}
\end{align}
Finally, subtracting Eq.~\eqref{eq:Helmholtz_add2} from Eq.~\eqref{eq:Helmholtz_add1} and making use of standard vector identities results in
\begin{align}
\frac{j}{2} \nabla \cdot &\left( \fE_p' \times \fH_q^* + \fE_q^* \times \fH_p' \right) \nonumber \\
=& \frac{1}{2} \left[ (1+\omega\frac{\varepsilon'}{\varepsilon}) \varepsilon \fE_q^* \cdot \fE_p + (1+\omega\frac{\mu'}{\mu}) \mu \fH_q^* \cdot \fH_p \right] \nonumber \\
&+ \frac{j}{2} \frac{\varepsilon - \varepsilon^*}{\varepsilon^*} \fE_p' \cdot \nabla \times \fH_q^*
- \frac{j}{2} \frac{\mu - \mu^*}{\mu^*} \fH_p' \cdot \nabla \times \fE_q^* \,.
\label{eq:Helmholtz_subtract1}
\end{align}
Integrating the left- and right-hand side (LHS and RHS) of Eq.~\eqref{eq:Helmholtz_subtract1} over $\Omega$ and applying the divergence theorem while accounting for the fact that electric fields tangential to all PEC walls vanish, yields
\begin{align}
&\frac{j}{2} \int_{\partial \Omega} \hat{\xi} \cdot \left( \fE_{p,\|}' \times \fH_{q,\|}^* + \fE_{q,\|}^* \times \fH_{p,\|}' \right) dS \nonumber \\
&\qquad \qquad \qquad = {Q}^e_{V,qp} + {Q}^d_{V,qp} + {Q}^l_{V,qp} \label{eq:Helmholtz_integ1}
\end{align}
where 
\begin{align}
&{Q}^e_{V,qp} = \frac{1}{2}  \int_{\Omega} \left( \varepsilon \fE_q^* \cdot \fE_p + \mu \fH_q^* \cdot \fH_p \right) dV \label{eq:Q_v_energy} \\
&{Q}^d_{V,qp} = \frac{1}{2}  \int_{\Omega} \left( \omega \varepsilon' \fE_q^* \cdot \fE_p + \omega \mu' \fH_q^* \cdot \fH_p \right) dV \label{eq:Q_v_disper}
\end{align}
\begin{align}
&{Q}^l_{V,qp} = \nonumber \\
& \quad \quad \frac{j}{2} \int_{\Omega} \Big( \frac{\varepsilon - \varepsilon^*}{\varepsilon^*} \fE_p' \cdot \nabla \times \fH_q^* - \frac{\mu - \mu^*}{\mu^*} \fH_p' \cdot \nabla \times \fE_q^* \Big) dV \,. \label{eq:Q_v_loss}
\end{align}
The subscripts \textit{V} suggest that the result is computed \textit{directly} by a volume integral and the superscripts $e$, $d$, $l$ indicate \textit{energy}, \textit{dispersion}, and \textit{loss}, respectively -- this terminology is further explained below. Eq.~\eqref{eq:Helmholtz_integ1} relates a surface integral of transverse field components across $\partial \Omega$ to volume integrals of fields across $\Omega$.

Taking the complex conjugate and frequency derivative of Eqs.~\eqref{eq:tot_E_transv} and \eqref{eq:tot_H_transv} and substituting the resulting expressions for $\fE_{q,\|}^*, \fH_{q,\|}^*$ and $\fE_{p,\|}', \fH_{p,\|}'$ into the LHS of Eq.~\eqref{eq:Helmholtz_integ1} allows the LHS of Eq.~\eqref{eq:Helmholtz_integ1} to be expressed as \cite{Patel2020WS1}
\begin{align}
& \frac{j}{2} \int_{\partial \Omega} \hat{\xi} \cdot \left( \fE_p' \times \fH_q^* + \fE_q^* \times \fH_p' \right) dS \nonumber \\
& \quad = j \sum_m S_{mq}^* S_{mp}'
- \frac{j}{2} S_{pq}^* \left( \frac{1}{Z_p} \right)' Z_p + \frac{j}{2} S_{qp} \left( \frac{1}{Z_q} \right)' Z_q \,. \label{eq:Helmholtz_integ1_LHS}
\end{align}
Upon defining
\begin{align}
Q_{V,qp}^b = Q_{V,qp}^e + \frac{j}{2} S_{pq}^* \left( \frac{1}{Z_p} \right)' Z_p - \frac{j}{2} S_{qp} \left( \frac{1}{Z_q} \right)' Z_q \,, \label{eq:Q_v_b}
\end{align}
where the superscript $b$ stands for \textit{base}, and
\begin{align}
Q_{V,qp} = Q_{V,qp}^b + Q_{V,qp}^d + Q_{V,qp}^l \,, \label{eq:Q_v_bdl}
\end{align}
it follows from Eqs.~\eqref{eq:Helmholtz_integ1} and \eqref{eq:Helmholtz_integ1_LHS} that
\begin{align}
Q_{V,qp} = j \sum_m S_{mq}^* S_{mp}' \,. \label{eq:WS_elem}
\end{align}
In matrix form, Eq.~\eqref{eq:WS_elem} reads
\begin{align}
\matr{Q}_{V} &= j \matr{S}^\dag \matr{S}' \,. \label{eq:WS_relation_1}
\end{align}
To arrive at Eq.~\eqref{eq:Q_def2}, define
\begin{align}
\matr{Q} &\coloneqq (\matr{S}^\dag \matr{S})^{-1} \matr{Q}_{V} \label{eq:Qv_to_Q} \\
&= j \matr{S}^{-1} \matr{S}' \label{eq:WS_relation_2} \,.
\end{align}

\subsection{Discussion}

The evaluation of the WS time delay matrix $\matr{Q}$ for lossy and dispersive systems involves the following steps.
\begin{enumerate}
    \item Evaluation of $\matr{Q}_V$, which consists of three contributions:
    \begin{enumerate}
         \item[1a)] Base contribution $\matr{Q}_V^b$, which is the sum of the energy integral $\matr{Q}_V^e$ in Eq.~\eqref{eq:Q_v_energy} and the impedance-related correction terms on the RHS of Eq.~\eqref{eq:Q_v_b}; note that the latter vanish for TEM modes.
         \item[1b)] Dispersion contribution $\matr{Q}_V^d$ in Eq.~\eqref{eq:Q_v_disper}, which vanishes when $\varepsilon' = \mu' = 0$.
         \item[1c)] Loss contribution $\matr{Q}_V^l$ in Eq.~\eqref{eq:Q_v_loss}, which vanishes when $\varepsilon = \varepsilon^*$, $\mu = \mu^*$.
     \end{enumerate}
    \item Evaluation of $\matr{Q}$ from $\matr{Q}_V$ via Eq.~\eqref{eq:Qv_to_Q}. This operation disentangles time delay and loss contributions inherent in $\matr{Q}_V$, and ensures that the elements of $\matr{Q}$ have clean time delay interpretations, as discussed in Section~\ref{sec:sys}.
\end{enumerate}

The contributions of dispersion and losses generalize expressions derived in \cite{Yagh2005imp} for single port antennas to a multi-port setting.

In the absence of loss and dispersion, $\matr{S}$ is unitary ($\matr{S}^\dag \matr{S} = \matr{I}$). It follows that $\matr{Q} = \matr{Q}_V = \matr{Q}_V^b$, and the above derivation reverts to the \textit{base} formulation applicable to lossless and dispersion-free systems in \cite{Patel2020WS1}.

The expression for $\matr{Q}_V^l$ in Eq.~\eqref{eq:Q_v_loss} involves $\fE_p'$ and $\fH_p'$, frequency derivatives of fields, seemingly increasing the computational cost of evaluating the WS time delay matrix for lossy systems. These quantities however can be easily evaluated via an approach that incurs only marginal additional computational cost, detailed in Appendix \ref{Appdix:freq_deri2}. With this approach, the computation of $\matr{Q}$ only involves the numerical solutions of Maxwell's equations at a single frequency and it remains convenient to obtain $\matr{S}'$ via $\matr{S}' = -j \matr{S} \matr{Q}$.

\section{Time Delay Interpretation}
\label{sec:sys}

\subsection{Time Delays}

The manipulation of Maxwell's equations and Gauss theorem produces Eqs.~\eqref{eq:WS_relation_1}--\eqref{eq:WS_relation_2} and facilitates the volume integral evaluation of $\matr{Q}$. However, $\matr{Q}$'s physical meaning and properties are not yet clear. To interpret $\matr{Q}$, consider the following setting.

For a linear, time-invariant and lossy system with only one port, assume a time-harmonic incoming signal
\begin{align}
E^i = e^{j \beta(\omega) \xi}
\end{align}
where $\xi$ is the distance from the port and is positive outside the system; $\beta(\omega)$ denotes the propagation constant (outside the system) at frequency $\omega$ and is purely real. Correspondingly, the system generates an outgoing signal
\begin{align}
E^o = S(\omega) e^{- j \beta(\omega) \xi}
\end{align}
where $S(\omega) = | S(\omega) | e^{- j \gamma(\omega)}$ is the scattering coefficient. $| S(\omega) | < 1$ due to the loss and $\gamma(\omega)$ represents the system's phase delay.

In order to characterize the system's group delay, consider a narrow-band incoming pulse signal with center frequency $\omega_0$, bandwidth $2 \Delta \omega$ and real-valued envelop $\widetilde{A}(t) = \int_{-\Delta \omega}^{\Delta \omega} A(\omega) e^{j \omega t} d\omega$, 
\begin{align}
\widetilde{E}^i(t) &= \text{Re} \left[ \IntOmega A(\omega - \omega_0) e^{j (\omega t + \beta \xi)} d\omega \right] \nonumber \\
&= \widetilde{A} (t + \beta'(\omega_0) \xi) \cos(\omega_0 t + \beta(\omega_0) \xi) \,.
\label{eq:NarrowIn}
\end{align}
The system generates a corresponding outgoing pulse
\begin{align}
\widetilde{E}^o(t) & = \text{Re} \left[ \IntOmega S(\omega) A(\omega - \omega_0) e^{j (\omega t - \beta \xi)} d\omega \right] \nonumber \\
&\cong | S(\omega_0) | \widetilde{A} (\overline{t}) \cos [ \omega_0 t - \beta(\omega_0) \xi - \gamma(\omega_0) ] \nonumber \\
& \quad + | S(\omega_0) |' \frac{d \widetilde{A} (\overline{t})}{d t} \sin [ \omega_0 t - \beta(\omega_0) \xi - \gamma(\omega_0) ]
\label{eq:NarrowOut1}
\end{align}
where Taylor expansions for $\beta$, $\gamma$, $| S(\omega) |$ at $\omega_0$ are made use of. For notational convenience, $\overline{\omega} = \omega - \omega_0$ and $\overline{t} = t - \beta'(\omega_0) \xi - \gamma'(\omega_0)$ are defined. It is worth mentioning that $| S(\omega_0) |'$ is generally nonzero even for nondispersive but lossy media and it results in the deformation of the original wave packet. However, $\widetilde{A}$ is smooth enough for a narrow-band signal and the second term on the right-hand side (RHS) of Eq.~\eqref{eq:NarrowOut1} can be neglected.

Comparing the first term on the RHS of Eq.~\eqref{eq:NarrowOut1} with Eq.~\eqref{eq:NarrowIn}, at $\xi = 0$, the time delay of the envelop turns out to be 
\begin{align}
\gamma'(\omega_0) &= \text{Re} \left[ j S(\omega_0)^{-1} S'(\omega_0) \right] \nonumber \\
&= \text{Re} \left[Q\right] \,.
\end{align}

The above discussion validates our choice of Eq.~\eqref{eq:Q_def2} as the definition of time delay matrix $\matr{Q}$. 
By contrast, an analogy of Eq.~\eqref{eq:Q_def1} is 
\begin{align}
\text{Re} \left[Q_V\right] &= \left[ j S(\omega_0)^* S'(\omega_0) \right] \nonumber \\
&= | S(\omega_0) |^2 \gamma'(\omega_0) \,,
\end{align}
representing the product of the fraction of exiting energy and the signal's group delay.

\subsection{Simultaneous Diagonalization}

For a nondispersive and lossless multiport system where $\matr{Q}$ and $\matr{S}$ are matrices, their properties are discussed in \cite{Patel2020WS1}. In the absence of loss, $\matr{S}$ is symmetric and unitary; $\matr{Q}$ is Hermitian, which follows from the symmetry of $\matr{S}$. The diagonalization of $\matr{Q}$, \ie $\matr{Q} = \matr{W} \overline{\matr{Q}} \matr{W}^\dag$, yields fully decoupled WS modes and well-defined time delays, given by columns of $\matr{W}$ and diagonal elements of $\overline{\matr{Q}}$, respectively \cite{Patel2020WS1}.

In the presence of loss, $\matr{S}$ is symmetric but not unitary; therefore $\matr{Q}$ is not Hermitian. 
Suppose $\matr{Q}$ is diagonalized by $\matr{W}$, \ie
\begin{align}
\matr{Q} = \matr{W} \overline{\matr{Q}} \matr{W}^{-1} \,. \label{eq:Q_diag}
\end{align}
Note that $\matr{W}^{-1} = \matr{W}^\dag$ would only hold for Hermitian $\matr{Q}$ in lossless systems. 
By defining
\begin{align}
\overline{\matr{S}} = \matr{W}^T \matr{S} \matr{W} \,, \label{eq:S_bar}
\end{align}
it can be verified that $\overline{\matr{S}}$ is symmetric. Below $\overline{\matr{S}}$ is also shown to be diagonal, \ie $\matr{W}$ realizes the Takagi decomposition of $\matr{S}$ \cite{Horn2012matrix}. By $\matr{Q}$'s definition Eq.~\eqref{eq:Q_def2},
\begin{align}
\matr{S}' &= - j \matr{S} \matr{Q} \nonumber \\
&= - j \left( \matr{W}^T \right)^{-1} \overline{\matr{S}} \matr{W}^{-1} \matr{W} \overline{\matr{Q}} \matr{W}^{-1} \nonumber \\
&= - j \left( \matr{W}^T \right)^{-1} \overline{\matr{S}}~ \overline{\matr{Q}} \matr{W}^{-1} \,. \label{eq:dw_S_diag1}
\end{align}
Since $\matr{S}'$ is symmetric, 
\begin{align}
\left( \matr{S}' \right)^T &= - j \left( \matr{W}^T \right)^{-1} \overline{\matr{Q}} ~\overline{\matr{S}} \matr{W}^{-1} = \matr{S}' \,. \label{eq:dw_S_T_diag1}
\end{align}
Eqs.~\eqref{eq:dw_S_diag1} and \eqref{eq:dw_S_T_diag1} suggest $\overline{\matr{S}}~\overline{\matr{Q}} = \overline{\matr{Q}} ~\overline{\matr{S}}$. Since $\overline{\matr{Q}}$ is diagonal, $\overline{\matr{S}}$ must be diagonal. Making use of Eq.~\eqref{eq:dw_S_diag1}, $\matr{S}'$ can be diagonalized as
\begin{align}
\overline{\matr{S}}' &\coloneqq \matr{W}^T \matr{S}' \matr{W} \nonumber \\
&= - j \overline{\matr{S}} ~\overline{\matr{Q}} \,,
\end{align}
where $\overline{\matr{S}}'$ is diagonal. Now Eq.~\eqref{eq:Q_def2} is completely diagonalized. The simultaneous diagonalization of the matrices $\matr{Q}$, $\matr{S}$, $\matr{S}'$ indicates that columns of $\matr{W}$ are so-called WS modes associated with well-defined group delays given by elements of the diagonal matrix $\overline{\matr{Q}}$.

It is worth mentioning that $\matr{Q}_V$ and $\matr{S}$ do not simultaneously diagonalize except for a lossless system. This can be shown using similar procedures as above.

\subsection{WS modes}

Denote $\matr{V} = \left( \matr{W}^T \right)^{-1}$ and Eq.~\eqref{eq:S_bar} is rewritten as
\begin{align}
\matr{V} \overline{\matr{S}} &= \matr{S} \matr{W} \,.
\end{align}
Consider an excitation with the incoming signal
\begin{align}
E_{\mathrm{WS},q}^i = \sum_p W_{pq} E_p^i \,,
\end{align}
which is define as the $q$-th incoming WS mode. The above excitation generates outgoing signal $E^o = \sum_k b_k \left( E_k^i \right)^*$ where $b_k = \sum_m S_{km } W_{mq} = (\matr{S} \matr{W})_{kq}$. Therefore
\begin{align}
E^o &= \sum_k (\matr{S} \matr{W})_{kq} \left( E_k^i \right)^* \nonumber \\
&= \overline{S}_{qq} \sum_k V_{kq} \left( E_k^i \right)^* \nonumber\\
&= \overline{S}_{qq} E_{\mathrm{WS},q}^o \label{eq:Eo_Sqq}
\end{align}
where
\begin{align}
E_{\mathrm{WS},q}^o = \sum_k V_{kq} \left( E_k^i \right)^* \label{eq:E_O_WS_q}
\end{align}
is defined as the $q$-th outgoing WS mode. One can observe that each incoming WS mode $E_{\mathrm{WS},q}^i$ excites only the corresponding outgoing WS mode $E_{\mathrm{WS},q}^o$ with an attenuation factor $\overline{S}_{qq}$. This suggests that WS modes are completely decoupled for lossy systems at a fixed frequency.

Finally, the field distribution of a WS mode is formally defined to be the total field excited by the corresponding incoming WS mode, \ie
\begin{align}
E_{\mathrm{WS},q} = E_{\mathrm{WS},q}^i + \overline{S}_{qq} E_{\mathrm{WS},q}^o \,.
\end{align}

The above results simplify to \cite{Patel2020WS1} for lossless systems by using $\matr{V} = \matr{W}^*$ and $\overline{\matr{S}}=\matr{I}$. As an example, Eq.~\eqref{eq:E_O_WS_q} simplifies to $E_{\mathrm{WS},q}^o = \left( E_{\mathrm{WS},q}^i \right)^*$.

\subsection{Frequency Stability of WS modes}

WS modes not only stay decoupled at a fixed frequency, but also exhibit minimal dispersion with respect to a frequency perturbation. This property is demonstrated below for a lossy system.

Assume that $\widehat{\matr{W}}$ and $\widehat{\matr{V}}$ are constant matrices defined as $\widehat{\matr{W}} \equiv \matr{W}(\omega_0)$, $\widehat{\matr{V}} \equiv \matr{V}(\omega_0)$; at a close enough \textit{perturbed} frequency $\omega = \omega_0 + \delta \omega$, define
\begin{align}
\widehat{\matr{Q}}(\omega) &= \widehat{\matr{W}}^{-1} \matr{Q}(\omega) \widehat{\matr{W}} \label{eq:Q_hat} \\
\widehat{\matr{S}}(\omega) &= \widehat{\matr{W}}^T \matr{S}(\omega) \widehat{\matr{W}} \,. \label{eq:S_hat}
\end{align}
Note that $\widehat{\matr{Q}}(\omega_0) = \overline{\matr{Q}}(\omega_0)$ and $\widehat{\matr{S}}(\omega_0) = \overline{\matr{S}}(\omega_0)$. Substituting Eqs.~\eqref{eq:Q_hat}--\eqref{eq:S_hat} into $\matr{Q}$'s definition Eq.~\eqref{eq:Q_def2} yields
\begin{align}
\widehat{\matr{S}}'(\omega) = - j \widehat{\matr{S}}(\omega) \widehat{\matr{Q}}(\omega) \,.
\end{align}
Then, applying a first-order forward Euler approximation to $\widehat{\matr{S}}'(\omega)$ yields 
\begin{align}
\widehat{\matr{S}}(\omega_0 + \delta \omega) &\cong \widehat{\matr{S}}(\omega_0) - j \delta \omega \widehat{\matr{S}}(\omega_0) \widehat{\matr{Q}}(\omega_0) \nonumber \\
&= \overline{\matr{S}} (\matr{I} - j \delta \omega \overline{\matr{Q}}) \nonumber \\
&\cong \overline{\matr{S}} e^{- j \delta \omega \overline{\matr{Q}}} \,.
\end{align}
A \textit{perturbed} incoming ``WS mode'' constructed by eigenvectors at $\omega_0$, 
\begin{align}
E_{\widehat{\mathrm{WS}},q}^i = \sum_p \widehat{W}_{pq} E_p^i (\omega_0 + \delta \omega) \,,
\end{align}
would excite outgoing signal
\begin{align}
E^o \cong \overline{S}_{qq} e^{- j \delta \omega \overline{Q}_{qq}} E_{\widehat{\mathrm{WS}},q}^o \,, \label{eq:E_o_disturb}
\end{align}
where $E_{\widehat{\mathrm{WS}},q}^o$ denotes the \textit{perturbed} outgoing ``WS mode'', defined as
\begin{align}
E_{\widehat{\mathrm{WS}},q}^o = \sum_p \widehat{V}_{pq} \left( E_p^i (\omega_0 + \delta \omega) \right)^* \,.
\end{align}
This implies that WS modes are frequency stable with minimal dispersion, \ie they react to a frequency perturbation primarily by collecting phase delays while remaining decoupled. The difference between WS modes at frequency $\omega_0$ and $\omega_0 + \delta \omega$ is much smaller compared to that between any other bases including waveguide modes.

The above results simplify to \cite{Patel2020WS1} for lossless systems by using $\widehat{\matr{V}} = \widehat{\matr{W}}^*$ and $\overline{\matr{S}}=\matr{I}$. For example, Eq.~\eqref{eq:E_o_disturb} simplifies to $E^o \cong e^{- j \delta \omega \overline{Q}_{qq}} \left( E_{\widehat{\mathrm{WS}},q}^i \right)^*$.

\section{Illustrative Examples}
\label{sec:example}

This section analytically and numerically validates WS relationships Eqs.~\eqref{eq:WS_relation_1} and \eqref{eq:WS_relation_2}, and illustrates the effects of loss and dispersion on the WS characterization of a system. Throughout, matrices $\matr{Q}_V$ and $\matr{Q}$ are constructed by volume integration of fields via Eqs.~\eqref{eq:Q_v_energy}--\eqref{eq:Q_v_loss}, \eqref{eq:Q_v_b}--\eqref{eq:Q_v_bdl} and \eqref{eq:Qv_to_Q}. Fields and scattering matrices are computed using a high-order mode expansion-based finite-element code \cite{Jin2015finite}. When needed, $\matr{S}'$ is computed using a finite-difference approximation with frequency step size $d\omega = 2 \times 10^{-7} \omega$. Numerical errors incurred in the computation of $\matr{Q}_{V}$ and $\matr{Q}$ are defined as $\text{err}(\matr{Q}_{V}) = \| \matr{Q}_{V} - j \matr{S}^\dag \matr{S}' \|_F / \| \matr{Q}_{V} \|_F$ and $\text{err}(\matr{Q}) = \| \matr{Q} - j \matr{S}^{-1} \matr{S}'  \|_F / \| \matr{Q} \|_F$ where $\| \cdot \|_F$ denotes the Frobenius norm.

\subsection{Waveguide Homogeneously Filled with Low-Loss Material}

To analytically validate the formalism from Section \ref{sec:em}, consider a PEC-terminated rectangular waveguide of length $L$ and transverse dimensions $a \times b$ that is filled with a homogeneous, nondispersive, and lossy material with complex permittivity $\varepsilon = \varepsilon_1 - j \varepsilon_2$ and real permeability $\mu$. The waveguide is excited by a unit-power TE$_{p=(m,n)}$ mode with cut-off wavenumber $k_c = \sqrt{(m \pi / a)^2 + (n \pi / b)^2}$ and mode profile
\begin{align}
\label{eq:TE_mode}
\bm{\mathcal{X}}_{p} = &\frac{2}{k_c \sqrt{ab}} \Bigg[ \hat{x} \frac{n \pi}{b} \cos \left( \frac{m \pi}{a}x \right) \sin \left( \frac{n \pi}{b}y \right) \nonumber \\ 
&- \hat{y} \frac{m \pi}{a} \sin \left( \frac{m \pi}{a}x \right) \cos \left( \frac{n \pi}{b}y \right) \Bigg] \,.
\end{align}
The wavelength, speed of light, and wavenumber inside the waveguide are $\lambda = \frac{1}{f \sqrt{\varepsilon_1 \mu}}$, $c = 1/\sqrt{\varepsilon_1 \mu}$, and $k = \omega \sqrt{\varepsilon \mu} = k_1 - j k_2$, respectively. To simplify expressions for $\matr{S}$, $\matr{Q}_V$ and $\matr{Q}$, it is assumed that $\varepsilon_2 \ll \varepsilon_1$, $\varepsilon_2' \ll \varepsilon_1'$, and $L \gg \lambda$, which imply that $k_2 \ll k_1$ and $k_2' \ll k_1'$. Below, ``$\cong$'' implies equality up to an error of $\mathcal{O}(k_2^2)$. The propagation constant for the $p$-th mode therefore is
\begin{align}
\beta_{p} &= \beta_{p,1} - j \beta_{p,2} \nonumber \\
&= \sqrt{k^2 - k_c^2} \nonumber \\
&= \sqrt{k_1^2 - k_c^2} - j\frac{k_2 k_1}{\sqrt{k_1^2 - k_c^2}} + \mathcal{O}(k_2^2) \nonumber \\
&\cong k_1 \cos \theta_p - j \frac{k_2}{\cos \theta_p} \,,
\end{align}
where $\beta_{p,1} \cong \sqrt{k_1^2 - k_c^2}$, $\beta_{p,2} \cong k_2 k_1 / \beta_{p,1}$ and $\cos \theta_p \equiv \beta_{p,1} / k_1$.

To verify and interpret WS relationships~\eqref{eq:WS_relation_1} and \eqref{eq:WS_relation_2}, note that the scattering matrix is diagonal with elements
\begin{align}
S_{pp} = - e^{-2j \beta_p L} = - e^{-2 \beta_{p,2} L} e^{-2j \beta_{p,1} L} \,.
\end{align}
The RHS of Eq.~\eqref{eq:WS_relation_1} therefore equates to
\begin{align}
j S_{pp}^* S_{pp}' \cong 2L e^{-4 \beta_{p,2} L} \frac{\sqrt{\mu \varepsilon_1}}{\cos \theta_p} \,. \label{eq:case1_WS1_RHS}
\end{align}

To evaluate the LHS of Eq.~\eqref{eq:WS_relation_1}, note that the total fields inside the waveguide can be expressed as
\begin{align}
\fE_p &= \sqrt{Z_p} \bm{\mathcal{X}}_{p} (e^{-j \beta_p z} + S_{pp}e^{j \beta_p z}) \label{eq:case1_E_total} \\
\fH_p &= \fH_{p,\|} + \fH_{p,z} \label{eq:case1_H_total}
\end{align}
where
\begin{align}
Z_p &= \sqrt{\frac{\mu}{\varepsilon_1}} \frac{k_1}{\beta_{p,1}} \\
\fH_{p,\|} &= \frac{1}{\sqrt{Z_p}} (\hat{z} \times \bm{\mathcal{X}}_{p}) (e^{-j \beta_p z} - S_{pp}e^{j \beta_p z}) \\
\fH_{p,z} &= - \hat{z} j \frac{k_c \sqrt{Z_p}}{\omega \mu} \frac{2}{\sqrt{ab}} \cos\left( \frac{m \pi}{a} x \right) \cos\left( \frac{n \pi}{b} y \right) \nonumber \\
& \quad \quad (e^{-j \beta_p z} + S_{pp}e^{j \beta_p z}) \,. \label{eq:case1_H_z}
\end{align}
Substituting Eqs.~\eqref{eq:case1_E_total}--\eqref{eq:case1_H_z} into Eqs.~\eqref{eq:Q_v_energy}--\eqref{eq:Q_v_loss} and using Eqs.~\eqref{eq:Q_v_b}--\eqref{eq:Q_v_bdl} yields 
\begin{align}
Q_{V,pp} & \cong e^{-4 \beta_{p,2} L} \frac{\sqrt{\mu \varepsilon_1}}{\cos \theta_p} 2L \,. \label{eq:example_Qv}
\end{align}
Comparing this result with Eq.~\eqref{eq:case1_WS1_RHS}, it follows that 
\begin{align}
Q_{V,pp} = j S_{pp}^* S_{pp}' \,.
\end{align}
Furthermore, using $S_{pp}^* S_{pp} = e^{-4 \beta_{p,2} L}$, it follows from Eq.~\eqref{eq:WS_relation_2} that 
\begin{align}
Q_{pp} \cong \frac{\sqrt{\mu \varepsilon_1}}{\cos \theta_p} 2L \cong j S_{pp}^{-1} S_{pp}' \,. \label{eq:example_WS_approx}
\end{align}

Eq.~\eqref{eq:example_WS_approx} confirms that $Q_{pp}$ is the time spent by a light ray traveling to and from the PEC termination at angle $\theta_p$ w.r.t. the waveguide axis. Comparing Eqs.~\eqref{eq:example_Qv} and \eqref{eq:example_WS_approx} shows that $Q_{V,pp}=Q_{pp} e^{-4 \beta_{p,2} L}$, implying $Q_{V,pp}$ is the product of time delays and losses.

\subsection{Waveguide Inhomogeneously Filled with a Weakly Lossy Absorber}

\begin{figure}[H]
\centering\includegraphics[width=6.0cm]{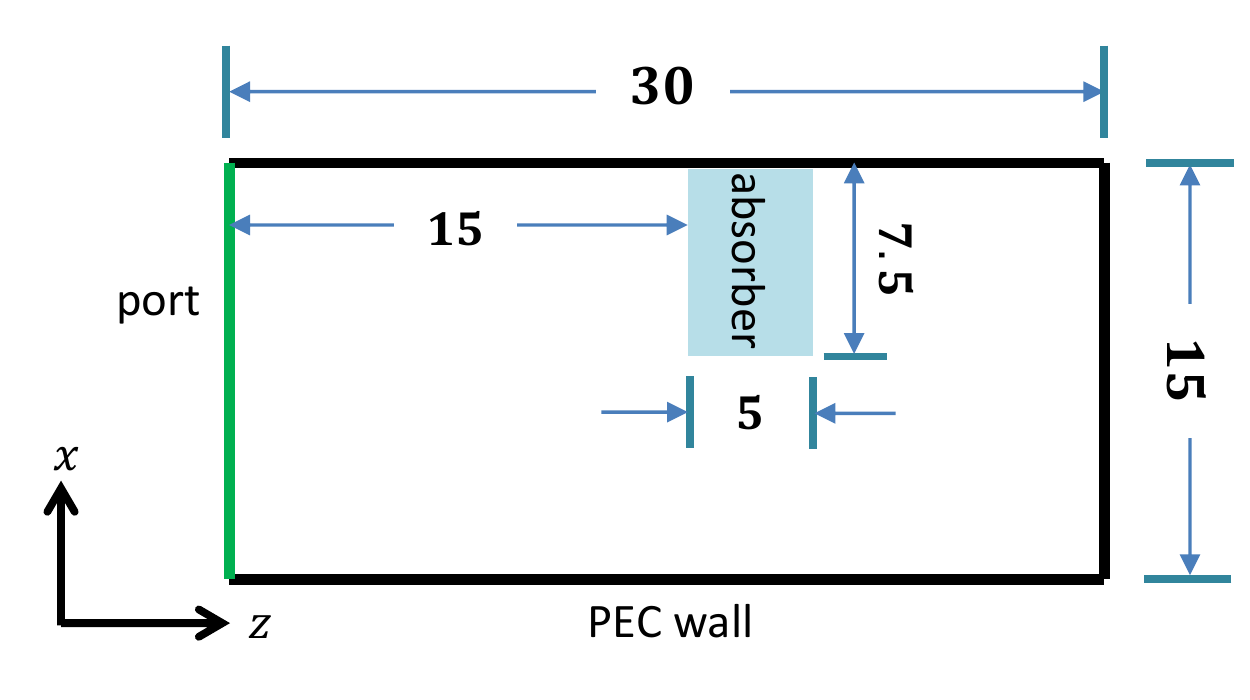}
\caption{PEC-terminated waveguide with an absorber (unit: mm). All structures extend infinitely in the $y$-direction.}
\label{fig:termin_inhomo}
\end{figure}

To investigate the similarities and differences between $\matr{Q}_V$ and $\matr{Q}$ for low-loss structures, consider the air-filled and PEC-terminated waveguide shown in Fig.~\ref{fig:termin_inhomo}, which contains a block of weakly absorbing dielectric material with relative permittivity $1.0 - 10^{-3}j$.  At $f=27.25$ GHz, the waveguide supports 27 propagating TE modes with profiles
\begin{align}
\bm{\mathcal{X}}_p = \sqrt{\frac{2}{a}} \sin{\left( \frac{p \pi x}{a} \right)} \hat{y} \,.
\end{align}

The skin depth in the absorber is 350 mm, significantly larger than the absorber's dimensions. The relative errors $\text{err}(\matr{Q}_V)$ and $\text{err}(\matr{Q})$ of the computed matrices $\matr{Q}_V$ and $\matr{Q}$ are $1.76 \times 10^{-5}$ and $1.85\times 10^{-5}$, respectively.

$\matr{Q}$ is diagonalized as $\matr{Q} = \matr{W} \overline{\matr{Q}} \matr{W}^{-1}$, and the real parts of its eigenvalues $\overline{\matr{Q}}$ are converted to ``spatial shifts'' by multiplying them with the speed of light in air. $\matr{Q}_V$ is diagonalized similarly as $\matr{Q}_V = \matr{W}_V \overline{\matr{Q}_V} \matr{W}_V^{-1}$. With some abuse of terminology, the products of the real parts of its eigenvalues $\overline{\matr{Q}}_V$ and the speed of light in air also are termed spatial shifts.

\begin{figure}[H]
\centering\includegraphics[width=7cm]{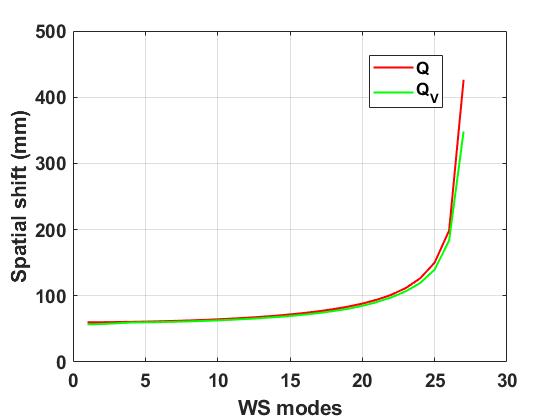}
\caption{Real parts of eigenvalues of $\matr{Q}$ and $\matr{Q}_V$ for the PEC-terminated waveguide with a weakly lossy absorber. The eigenvalues in time have been converted to equivalent spatial shifts by multiplying with the speed of light in free space.}
\label{fig:termin_inhomo_eig}
\end{figure}

\begin{figure}[H]
\null \hfill
\subfloat[$\matr{Q}$, mode \#1, 60.04 mm, \newline \hspace*{1.5em}$ \left| \overline{S}_{1,1} \right| = 0.9839$ \label{fig:inhomo_Q_WS1}]{\includegraphics[width=0.5\columnwidth]{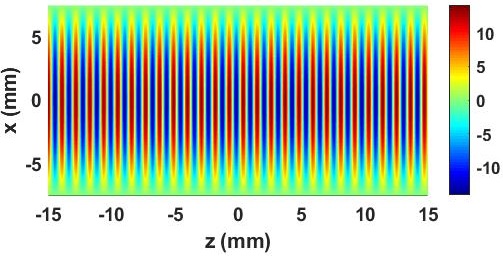}}
\null \hfill
\subfloat[$\matr{Q}$, mode \#2, 60.16 mm, \newline \hspace*{1.5em}$ \left| \overline{S}_{2,2} \right| = 0.9838$ \label{fig:inhomo_Q_WS2}]{\includegraphics[width=0.5\columnwidth]{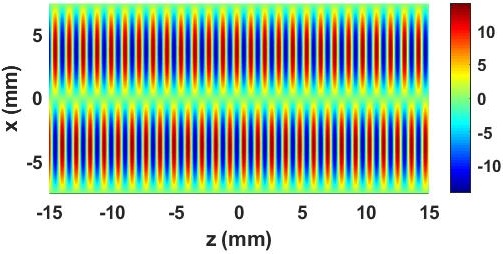}} 
\hfill \null \\
\null \hfill
\subfloat[$\matr{Q}_V$, mode \#1, 56.83 mm \label{fig:inhomo_Qv_WS1}]{\includegraphics[width=0.5\columnwidth]{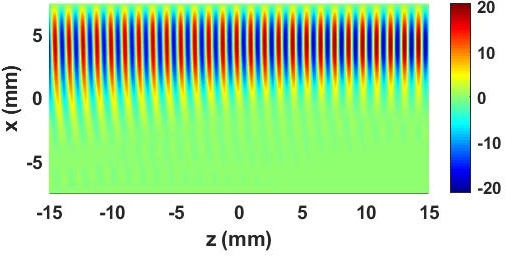}} \hfill
\subfloat[$\matr{Q}_V$, mode \#2, 57.31 mm \label{fig:inhomo_Qv_WS2}]{\includegraphics[width=0.5\columnwidth]{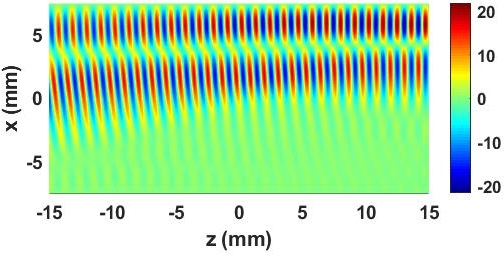}}
\hfill \null
\caption{$real(E_y)$ of modes obtained from $\matr{Q}$ and $\matr{Q}_V$, for the PEC-terminated waveguide with a weakly lossy absorber. Distance values below each subfigures are the corresponding spatial shifts in Fig.~\ref{fig:termin_inhomo_eig}.}
\label{fig:termin_inhomo_modes}
\end{figure}

Fig.~\ref{fig:termin_inhomo_eig} shows that spatial shifts obtained from $\matr{Q}$ and $\matr{Q}_V$ largely coincide. The similarity in the spatial shifts does not extend to the eigenvectors $\matr{W}$ (WS modes) and $\matr{W}_V$, however. To illustrate this, consider WS mode \#1, shown in Fig.~\ref{fig:inhomo_Q_WS1}. This mode experiences a spatial shift of 60.04 mm, which maps onto a round-trip between the port and waveguide termination. In contrast, the field associated with the first eigenvector of $\matr{Q}_V$, shown in Fig.~\ref{fig:inhomo_Qv_WS1}, is characterized by a spatial shift of 56.83 mm. This spatial shift and the field profile reflect the fact that the eigenvectors of $\matr{Q}_V$ minimize time delays while maximizing losses by preferentially traveling through the absorber block in the upper part of the waveguide. Similar observations hold true for WS mode \#2 and the field derived from $\matr{Q}_V$'s second eigenvector, shown in Figs.~\ref{fig:inhomo_Q_WS2} and \ref{fig:inhomo_Qv_WS2}, respectively. 
The $|\bar{S}_{qq}|$ values below Figs.~\ref{fig:inhomo_Q_WS1} and \ref{fig:inhomo_Q_WS2} are a measure of the attenuation that the WS modes experience; for this low-loss system, the magnitudes of all scattering parameters are close to one. 

\subsection{Waveguide Inhomogeneously Filled with a Strongly Lossy Absorber}

\begin{figure}[hbtp!]
\centering\includegraphics[width=6.0cm]{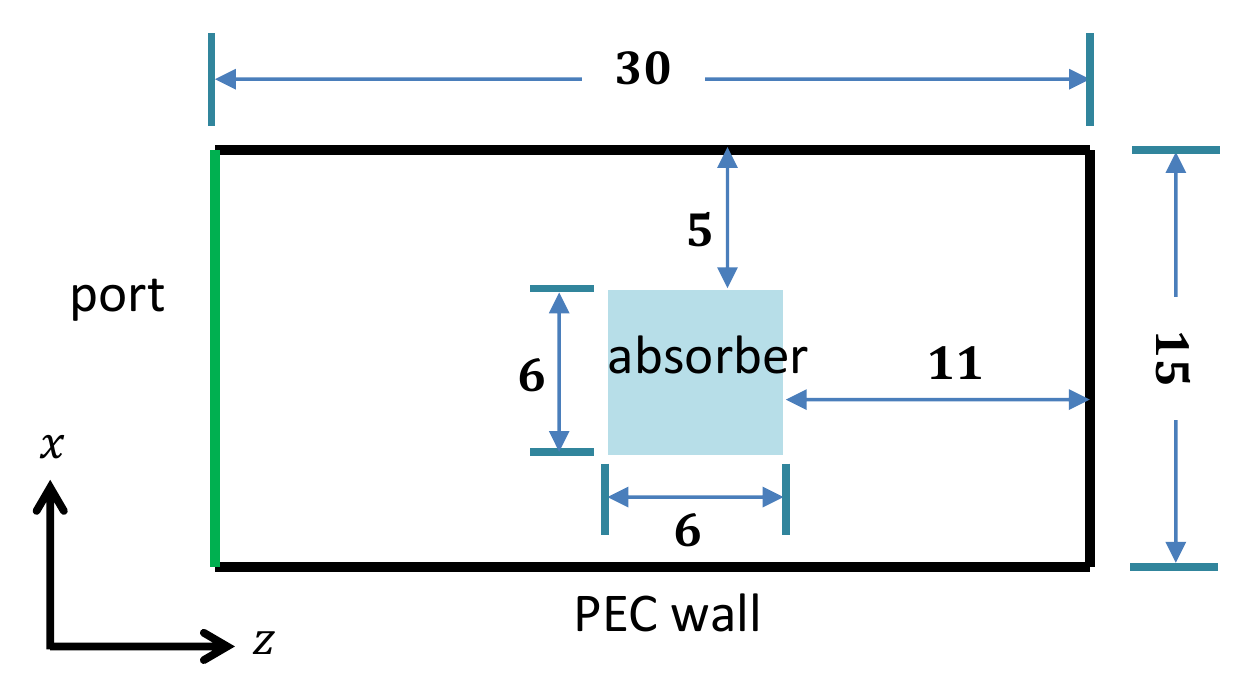}
\caption{PEC-terminated waveguide with a strongly lossy absorber (unit: mm). All structures extend infinitely in the $y$-direction.}
\label{fig:strong_loss}
\end{figure}

To investigate differences between $\matr{Q}_V$ and $\matr{Q}$ for highly lossy structures, consider the air-filled and PEC-terminated waveguide shown in Fig.~\ref{fig:strong_loss}. The waveguide and its excitation are identical to those in the previous section, except for the position and material properties of the lossy block, which now has a relative permittivity of $1.0 - 0.05 j$ and a skin depth of 7.0 mm. The relative errors $\text{err}(\matr{Q}_V)$ and $\text{err}(\matr{Q})$ of the computed matrices $\matr{Q}_V$ and $\matr{Q}$ once again are $8.6 \times 10^{-6}$ and $7.8 \times 10^{-5}$, respectively.

\begin{figure}[H]
\centering\includegraphics[width=7.0cm]{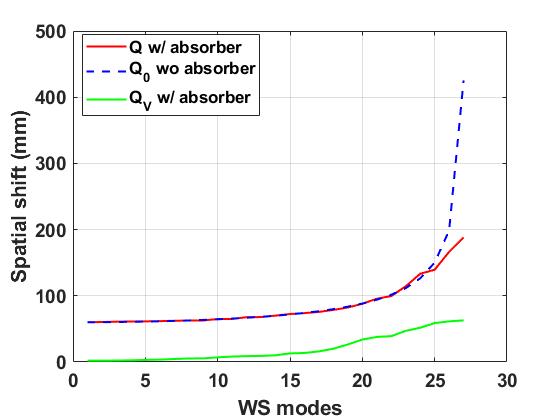}
\caption{Spatial shifts derived from the real parts of the eigenvalues of $\matr{Q}$ and $\matr{Q}_V$ for the PEC-terminated waveguide of Fig.~\ref{fig:strong_loss} loaded with a high-loss absorber; also shown are the spatial shifts obtained from $\matr{Q}_0$ for the empty waveguide.}
\label{fig:strong_loss_time_delay}
\end{figure}

Spatial shifts obtained by diagonalizing $\matr{Q}$ and $\matr{Q}_V$ are shown in Fig.~\ref{fig:strong_loss_time_delay}.  Also shown are spatial shifts obtained by diagonalizing WS time delay matrix $\matr{Q}_0$ for the waveguide without the absorber. Clearly, the absorber does not noticeably affect the time delays of most WS modes relative to those observed in the empty waveguide, as the real part of its permittivity is identical to that of the surrounding air. The eigenvalues of $\matr{Q}_V$, instead, no longer match those of $\matr{Q}$ as was the case for the low-loss structure. This fact can be ascribed wholesale to the large material losses.

\begin{figure}[H]
\null \hfill
\subfloat[mode \#1, 60.21 mm, \newline \hspace*{1.5em}$ \left| \overline{S}_{1,1} \right| = 0.6620$ \label{fig:strong_loss_WS1}]{\includegraphics[width=0.5\columnwidth]{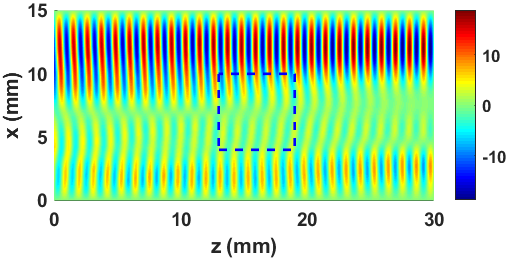}}
\null \hfill
\subfloat[mode \#6, 61.71 mm, \newline \hspace*{1.5em}$ \left| \overline{S}_{6,6} \right| = 0.4643$ \label{fig:strong_loss_WS6}]{\includegraphics[width=0.5\columnwidth]{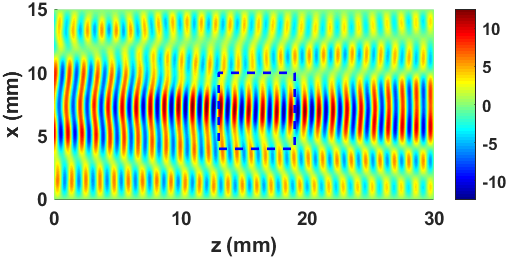}} 
\hfill \null \\
\null \hfill
\subfloat[mode \#15, 72.48 mm, \newline \hspace*{1.5em}$ \left| \overline{S}_{15,15} \right| = 0.3111$ \label{fig:strong_loss_WS15}]{\includegraphics[width=0.5\columnwidth]{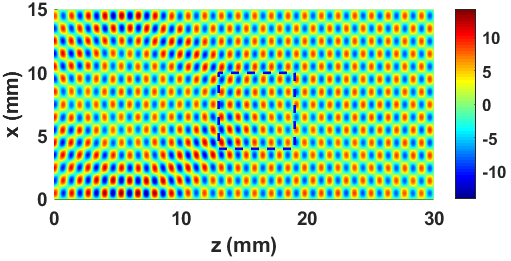}} \hfill
\subfloat[mode \#27, 188.2 mm, \newline \hspace*{1.5em}$ \left| \overline{S}_{27,27} \right| = 0.1661$ \label{fig:strong_loss_WS27}]{\includegraphics[width=0.5\columnwidth]{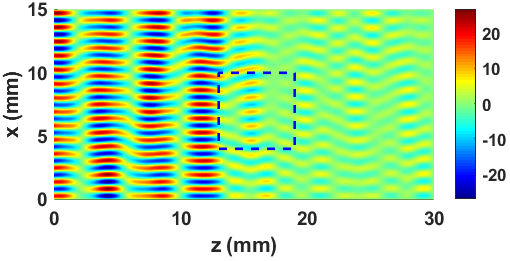}}
\hfill \null
\caption{$real(E_y)$ of selected WS modes for the PEC-terminated waveguide with a strongly lossy absorber. Distance values below each subfigures are the corresponding spatial shifts in Fig.~\ref{fig:strong_loss_time_delay}.}
\label{fig:strong_loss_modes}
\end{figure}

Fig.~\ref{fig:strong_loss_modes} shows several WS modes constructed from the eigenvectors of $\matr{Q}$. 
\begin{enumerate}
\item WS modes \#1 and \#6 are fields that primarily reflect off the waveguide termination instead of the waveguide's sidewalls, and experience very similar spatial shifts equating to roughly twice the waveguide length. Despite the similarity in spatial shifts, these modes' $|\bar{S}_{qq}|$'s differ substantially because mode \#1 involves waves that mostly bypass the material block whereas those of mode \#6 pass straight through it. 
\item WS mode \#15 travels along a zigzag pattern, reflecting off both the termination and the PEC walls, resulting in a larger spatial shift. The $|\bar{S}_{15,15}|$ value is smaller compared to that of mode \#6, as mode \#15 excites waves traveling diagonally through the absorbing block and suffering from bigger losses. 
\item WS mode \#27 excites waves that travel almost parallel to the port surfaces. This waves travel a much longer distance than Ws mode \#6 prior to exiting via the port, thus experiencing the longest time delay among all WS modes. 
Both the field distribution and the small $|\bar{S}_{27,27}|$ indicate this modes experiences significant attenuation. That said, Fig.~\ref{fig:strong_loss_time_delay} shows that mode \#27's time delay (red) is smaller than that of the corresponding mode in the absence of the absorber (dashed blue), suggesting that the presence of the absorber causes reflections limiting field penetration deep into the waveguide. 
\end{enumerate}

\subsection{Waveguide with Material with Anomalous Dispersion}

\begin{figure}[H]
\centering\includegraphics[width=7.0cm]{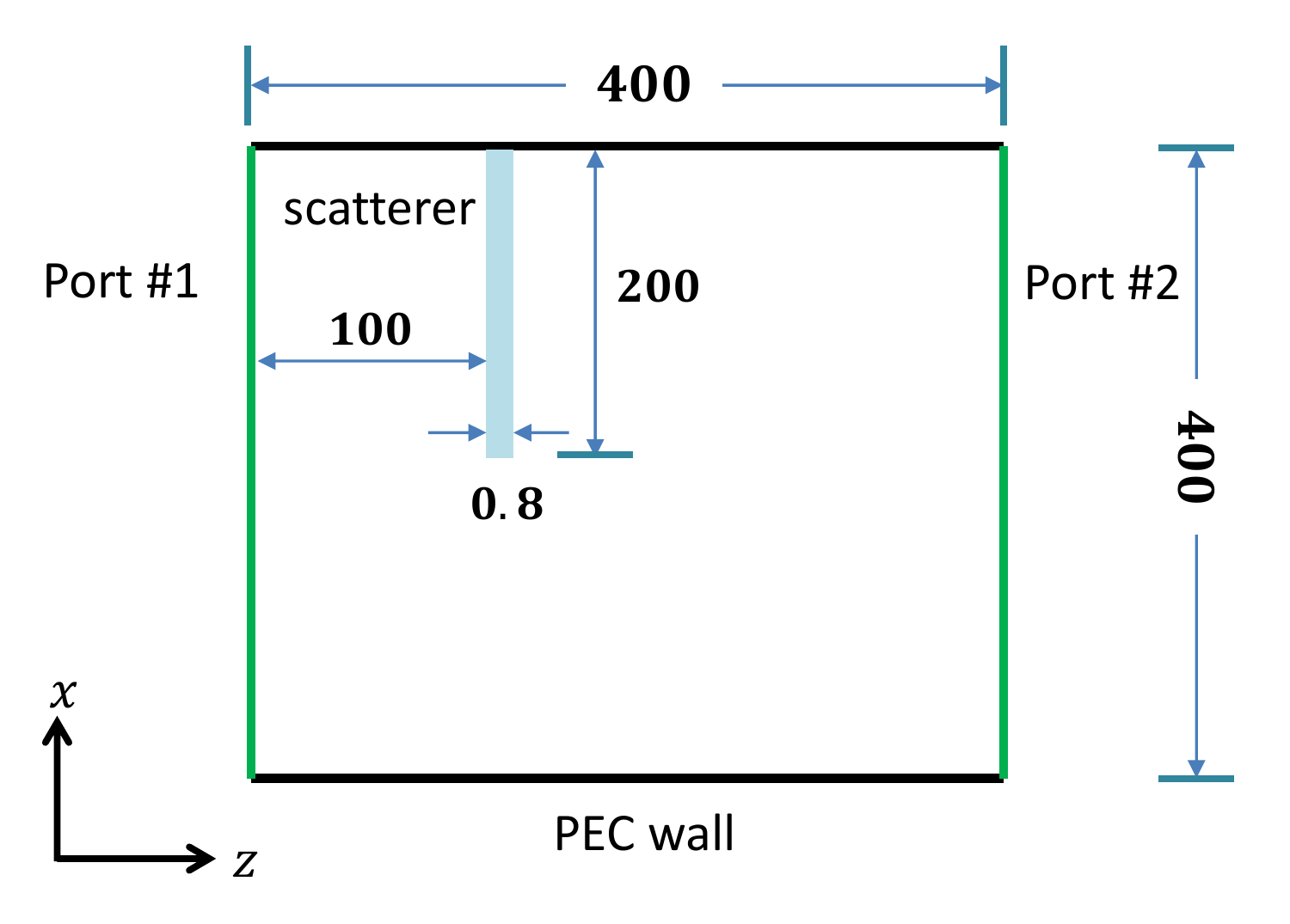}
\caption{Two-port waveguide with a dispersive absorber in its anomalous dispersion band (unit: $\mu$m). All structures extend infinitely in the $y$-direction.}
\label{fig:disper}
\end{figure}

To illustrate the effect of dispersion on WS modes, consider the (somewhat artificial) two-port air-filled waveguide containing a dispersive NaCl insert shown in Fig.~\ref{fig:disper}. At frequency $f = 5.081 \times 10^{3}$ GHz, NaCl exhibits anomalous dispersion; its relative permittivity is $\varepsilon_r = 3.4644 - 18.506j + d \varepsilon_r / d f (f - f_0)$ and $d \varepsilon_r / d f = -8.0701 \times 10^{-5} + 3.4834 \times 10^{-5} j$ Hz$^{-1}$ \cite{RefracInfo, Querry1987optical}.

\begin{figure}[H]
\centering\includegraphics[width=7.0cm]{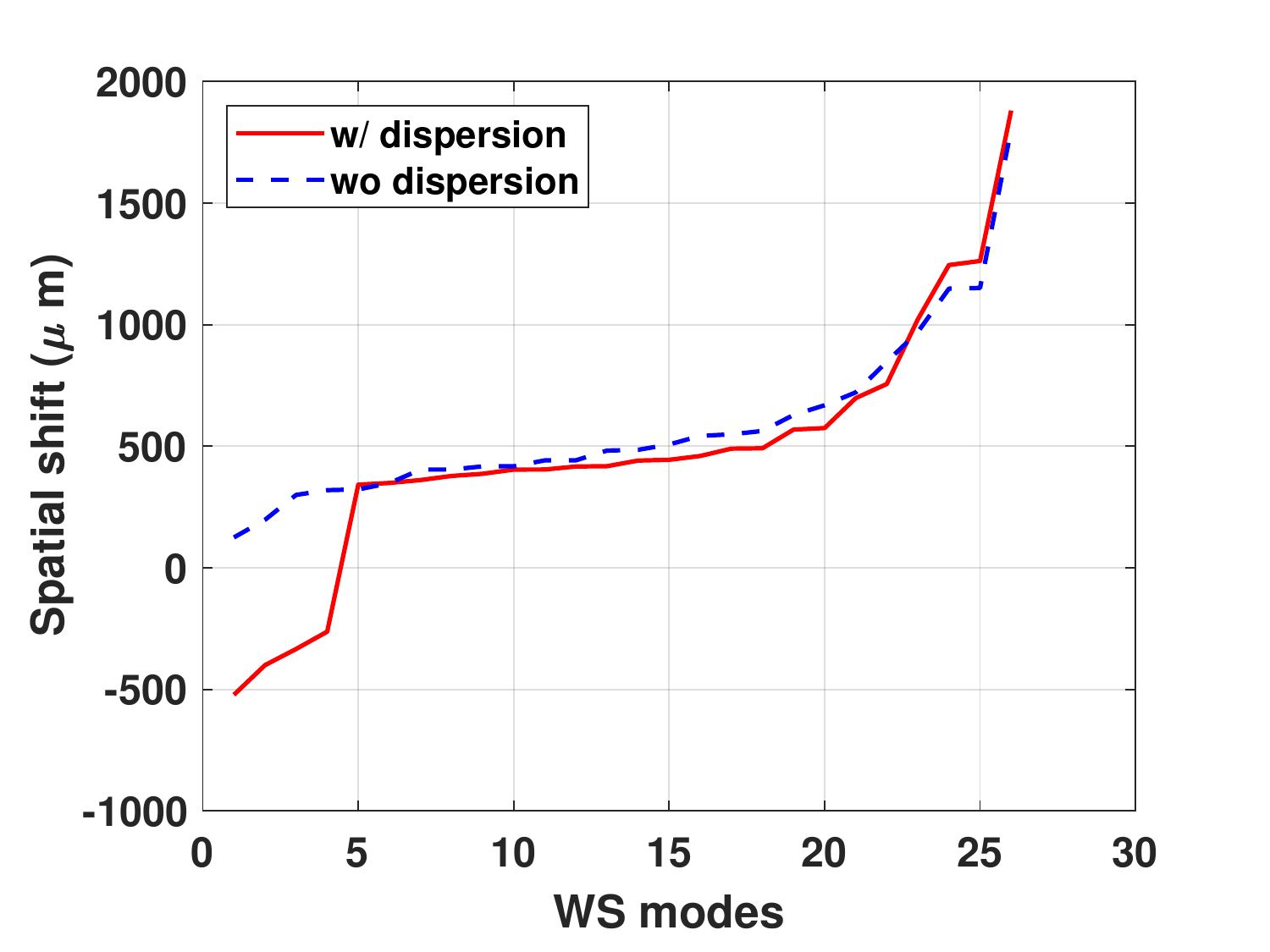}
\caption{Time delays converted to equivalent spatial shift by multiplying the real part of $\matr{Q}$'s eigenvalues with the speed of light in air, for the air-filled two-port waveguide in Fig.~\ref{fig:disper}. The red and dashed blue curves are obtained with and without dispersion, respectively.}
\label{fig:disper_eig}
\end{figure}

The matrix $\matr{Q}$ is constructed with numerical error $\text{err}(\matr{Q}) = 1.7 \times 10^{-4}$. Spatial shifts obtained upon diagonalizing $\matr{Q}$, are shown in Fig.~\ref{fig:disper_eig}; also shown are spatial shifts assuming the material insert has no dispersion, \ie $d \varepsilon_r / d f = 0$. The first four time delays for the system with dispersion are negative, indicating negative group delays, while all the time delays for the system without dispersion are positive. The negative group delays can be attributed to the anomalous dispersion. 
Indeed, for the system with dispersion, the overall group delay experienced by a WS mode is a combination of positive and negative group delays as the wave travels outside and inside the NaCl insert. 

\begin{figure}[htbp!]
\null \hfill
\subfloat[mode \#1, -524.1 $\mu$m, \newline \hspace*{1.5em}$ \left| \overline{S}_{1,1} \right| = 0.0823$ \label{fig:disper_WS1}]{\includegraphics[width=0.45\columnwidth]{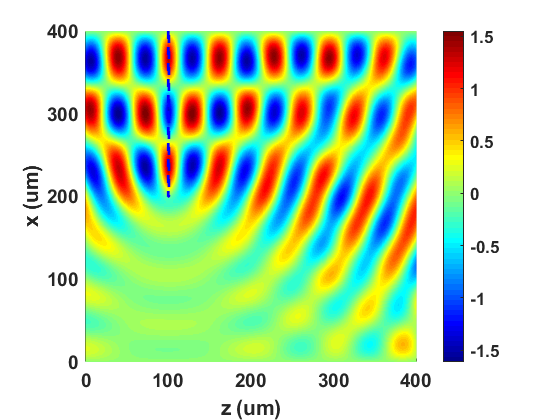}} \hfill
\subfloat[mode \#3, -333.9 $\mu$m, \newline \hspace*{1.5em}$\left| \overline{S}_{3,3} \right| = 0.1207$ \label{fig:disper_WS3}]{\includegraphics[width=0.45\columnwidth]{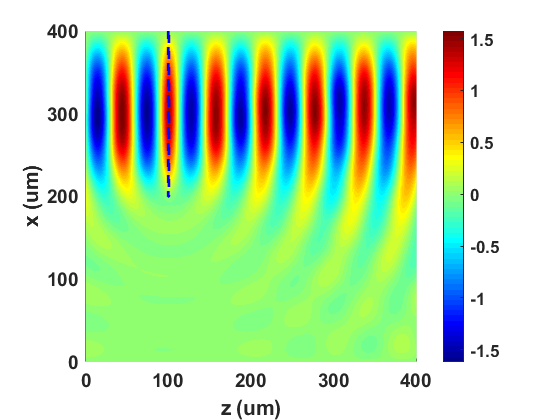}} \hfill
\hfill \null \\
\null \hfill
\subfloat[mode \#10, 404.0 $\mu$m, \newline \hspace*{1.5em}$\left| \overline{S}_{10,10}  \right| = 0.9041$ \label{fig:disper_WS10}]{\includegraphics[width=0.45\columnwidth]{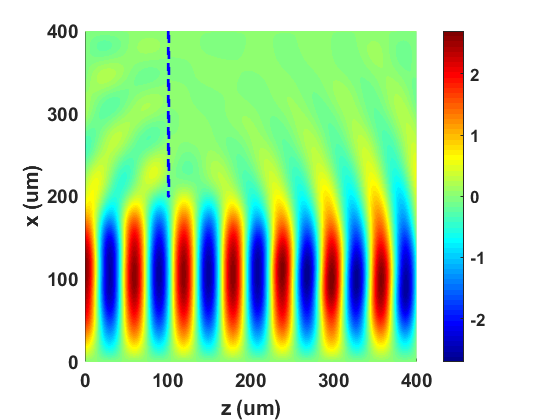}} \hfill
\subfloat[mode \#17, 490.3 $\mu$m, \newline \hspace*{1.5em}$\left| \overline{S}_{17,17} \right| = 0.8111$ \label{fig:disper_WS17}]{\includegraphics[width=0.45\columnwidth]{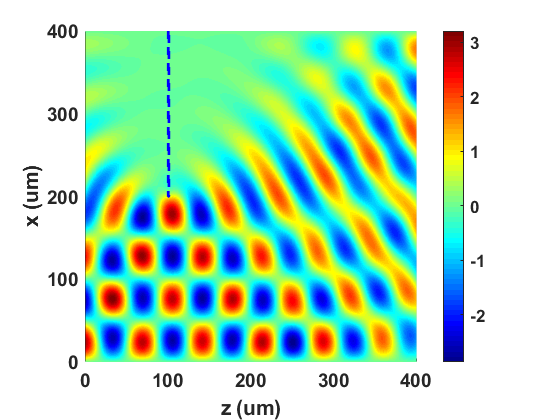}} \hfill
\caption{$real(E_y)$ of several WS modes for the waveguide with a dispersive scatterer. The dashed line indicates the outline of the scatterer. Distance values below each subfigures are the corresponding spatial shifts in Fig.~\ref{fig:disper_eig}.}
\label{fig:disper_modes}
\end{figure}

Several representative WS modes are shown in Fig.~\ref{fig:disper_modes}.
The most negative group delay is experienced by WS mode \#1 (Fig.~\ref{fig:disper_WS1}), which travels through the NaCl insert obliquely to maximize its interaction with the latter. WS mode \#3 also exhibits a negative group delay with a smaller magnitude by traveling nearly normal to the surface of the insert (Fig.~\ref{fig:disper_WS3}). 
These two modes have very small $|\bar{S}_{qq}|$ as their energies are effectively absorbed by the insert. 
In contrast, WS modes \#10 and \#17 (Figs.~\ref{fig:disper_WS10} and \ref{fig:disper_WS17}) experience positive group delays, tend to avoid interactions with the insert, and have large $|\bar{S}_{qq}|$. 

Therefore, due to the impact of dispersion, the group delays differ from time delays measured by phase propagation. The difference can be significant if $\omega \varepsilon'$ and $\omega \mu'$ in Eq.~\eqref{eq:Q_v_disper} are large compared to $\varepsilon$ and $\mu$ in Eq.~\eqref{eq:Q_v_energy}.

\subsection{Three-port Waveguide with GaAs Material}

\begin{figure}[hbtp!]
\centering\includegraphics[width=7.0cm]{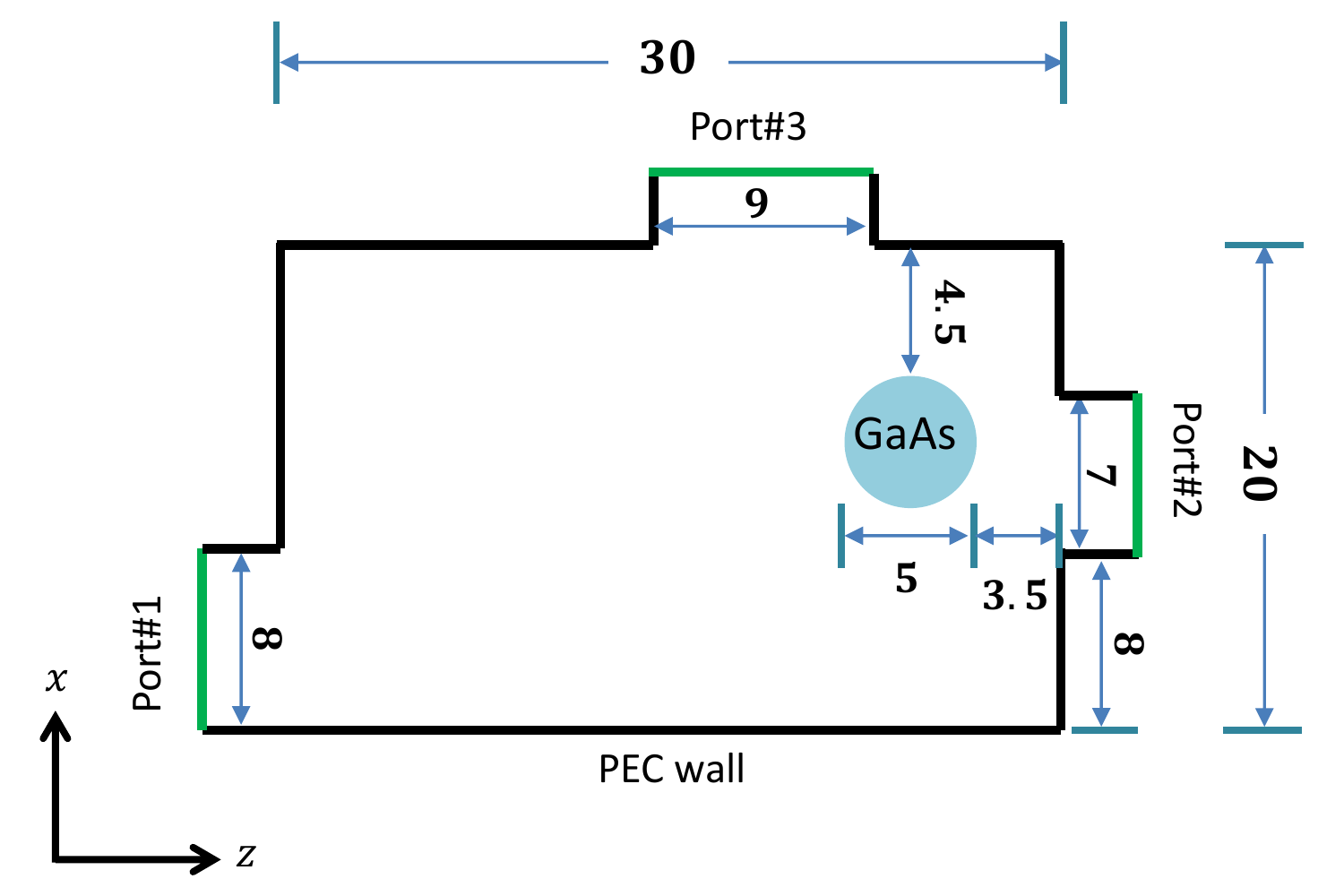}
\caption{Three-port waveguide with a cylindrical scatterer (unit: $\mu$m). All structures extend infinitely in the $y$-direction.}
\label{fig:threeport}
\end{figure}

To illustrate the use of the methods of Section II on a slightly more complicated geometry, consider the three-port air-filled waveguide loaded with a dispersive and lossy GaAs cylinder shown in Fig.~\ref{fig:threeport}. At $f_0 = 3.89 \times 10^5$ GHz, GaAs has a relative permittivity $\varepsilon_r = 13.417 - 0.656j + d \varepsilon_r / d f (f - f_0)$, and $d \varepsilon_r / d f = 7.9105 \times 10^{-15} - 7.0435 \times 10^{-15} j$ Hz$^{-1}$ \cite{RefracInfo, Papa2021Refrac}.

\begin{figure}[H]
\centering\includegraphics[width=7.0cm]{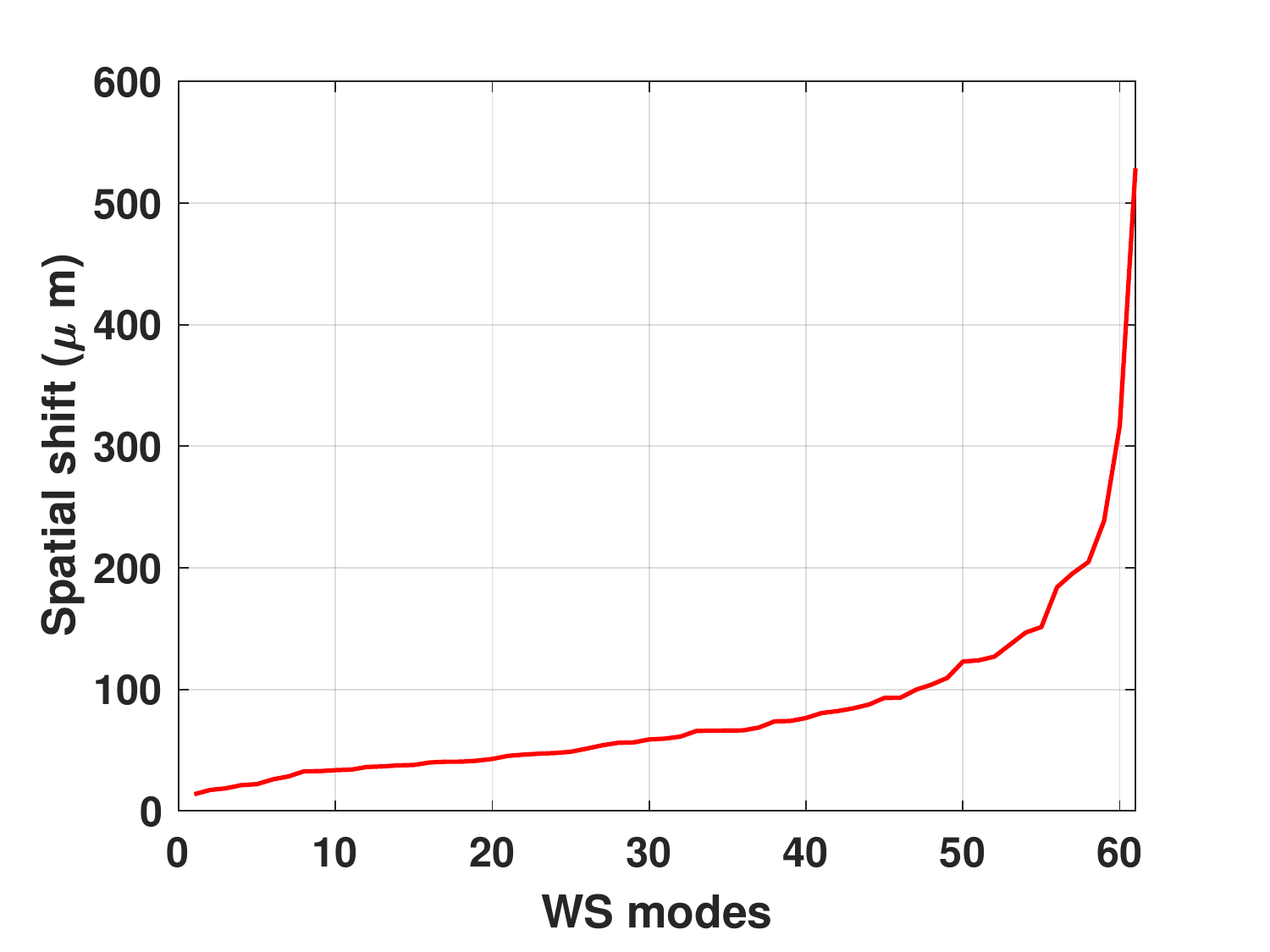}
\caption{Time delays converted to equivalent spatial shift by multiplying the real part of $\matr{Q}$'s eigenvalues with the speed of light in air, for the three-port waveguide with a GaAs cylinder.}
\label{fig:port3_circ_eig}
\end{figure}

\begin{figure*}[htbp!]
\null \hfill
\subfloat[mode \#1, 13.76 $\mu$m, \newline \hspace*{1.1em}
$ \left| \overline{S}_{1,1} \right| = 0.5746$ \label{fig:port3_Q_WS1}]{\includegraphics[width=0.5\columnwidth]{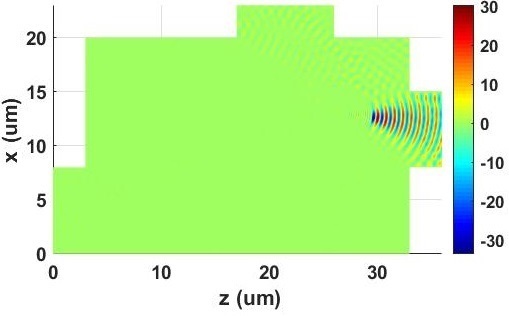}} \hfill
\subfloat[mode \#3, 18.61 $\mu$m, \newline \hspace*{1.1em}
$ \left| \overline{S}_{3,3} \right| = 0.5330$ \label{fig:port3_Q_WS3}]{\includegraphics[width=0.5\columnwidth]{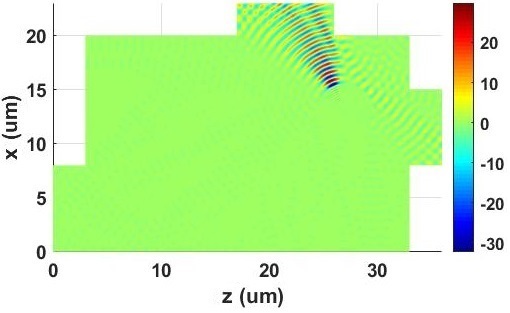}} \hfill
\subfloat[mode \#4, 21.23 $\mu$m, \newline \hspace*{1.1em}
$ \left| \overline{S}_{4,4} \right| = 0.8094$ \label{fig:port3_Q_WS4}]{\includegraphics[width=0.5\columnwidth]{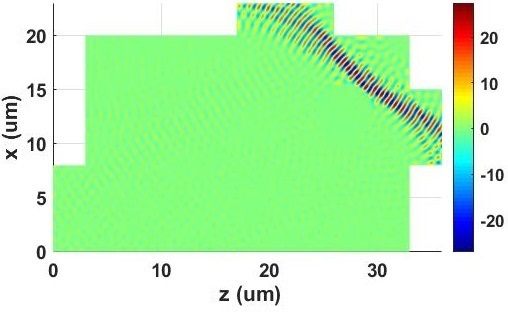}} \hfill
\subfloat[mode \#8, 32.64 $\mu$m, \newline \hspace*{1.1em}
$ \left| \overline{S}_{8,8} \right| = 0.4795$ \label{fig:port3_Q_WS8}]{\includegraphics[width=0.5\columnwidth]{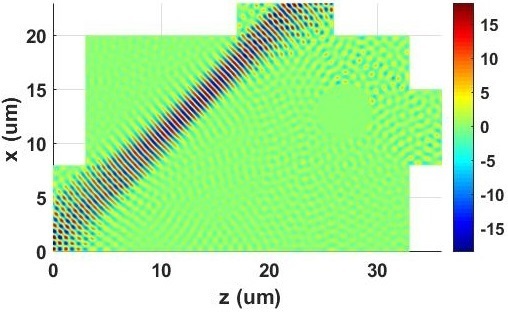}}
\hfill \null \\
\null \hfill
\subfloat[mode \#13, 36.66 $\mu$m, \newline \hspace*{1.1em}
$ \left| \overline{S}_{13,13} \right| = 0.9502$ \label{fig:port3_Q_WS13}]{\includegraphics[width=0.5\columnwidth]{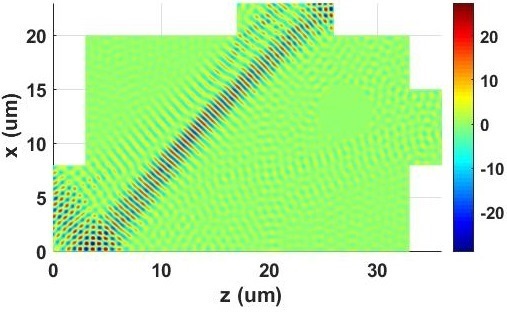}} \hfill
\subfloat[mode \#14, 37.50 $\mu$m, \newline \hspace*{1.1em}
$ \left| \overline{S}_{14,14} \right| = 0.9644$ \label{fig:port3_Q_WS14}]{\includegraphics[width=0.5\columnwidth]{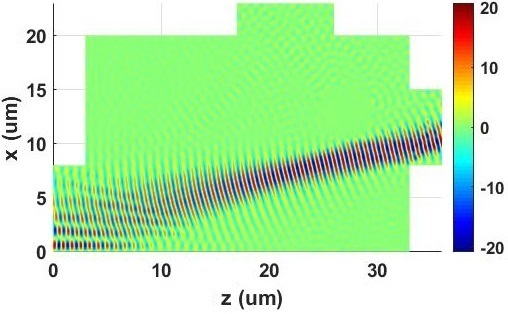}} \hfill
\subfloat[mode \#16, 39.85 $\mu$m, \newline \hspace*{1.1em}
$ \left| \overline{S}_{16,16} \right| = 0.8920$ \label{fig:port3_Q_WS16}]{\includegraphics[width=0.5\columnwidth]{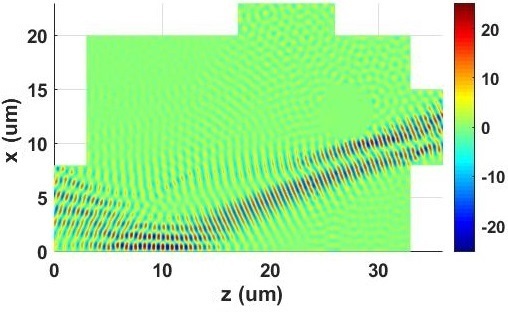}} \hfill
\subfloat[mode \#22, 46.36 $\mu$m, \newline \hspace*{1.1em}
$ \left| \overline{S}_{22,22} \right| = 0.8419$ \label{fig:port3_Q_WS22}]{\includegraphics[width=0.5\columnwidth]{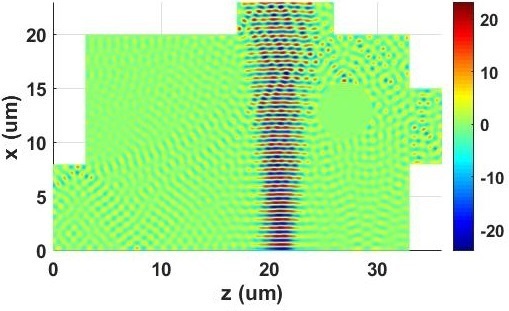}}
\hfill \null \\
\null \hfill
\subfloat[mode \#36, 66.31 $\mu$m, \newline \hspace*{1.1em}
$ \left| \overline{S}_{36,36} \right| = 0.8099$ \label{fig:port3_Q_WS36}]{\includegraphics[width=0.5\columnwidth]{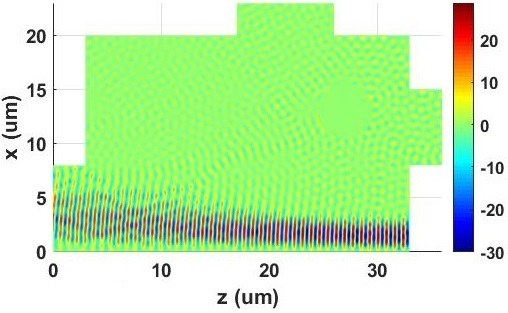}} \hfill
\subfloat[mode \#42, 82.13 $\mu$m, \newline \hspace*{1.1em}
$ \left| \overline{S}_{42,42} \right| = 0.5301$ \label{fig:port3_Q_WS42}]{\includegraphics[width=0.5\columnwidth]{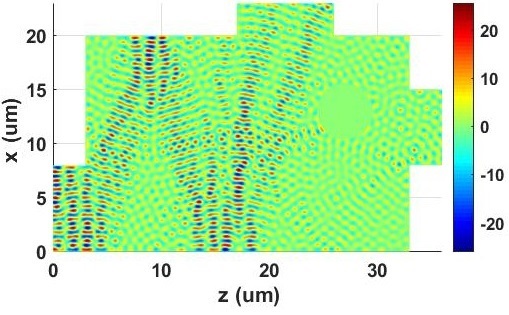}} \hfill
\subfloat[mode \#56, 184.1 $\mu$m, \newline \hspace*{1.1em}
$ \left| \overline{S}_{56,56} \right| = 0.5280$ \label{fig:port3_Q_WS56}]{\includegraphics[width=0.5\columnwidth]{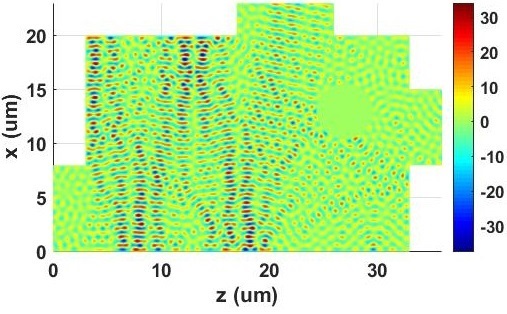}} \hfill
\subfloat[mode \#61, 528.8 $\mu$m, \newline \hspace*{1.1em}
$ \left| \overline{S}_{61,61} \right| = 0.1024$ \label{fig:port3_Q_WS61}]{\includegraphics[width=0.5\columnwidth]{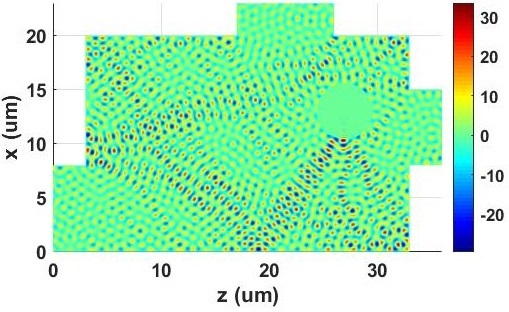}}
\caption{$real(E_y)$ of several WS modes for the three-port waveguide with a GaAs cylinder. Distance values below each subfigures are the corresponding spatial shifts in Fig.~\ref{fig:port3_circ_eig}.}
\label{fig:port3_circ_modes}
\end{figure*}

Ports \#1, \#2, and \#3 support 20, 23, and 18 propagating TE modes, respectively; $\matr{S}$ and $\matr{Q}$ therefore both are 61-by-61. The relative errors of the matrices $\matr{Q}_V$ and $\matr{Q}$ are $\text{err}(\matr{Q}_V) = 7.7 \times 10^{-4}$ and $\text{err}(\matr{Q}) = 6.3 \times 10^{-3}$, respectively. Diagonalization of $\matr{Q}$ yields the spatial shifts and WS mode field distributions shown in Figs.~\ref{fig:port3_circ_eig} and ~\ref{fig:port3_circ_modes}, respectively. These WS modes are categorized as follows.

\begin{enumerate}
\item WS modes \#1 and \#3 excite waves that exit the system immediately once reflected by the scatterer, experiencing very short time delays. 
\item WS modes \#4, \#8 and \#14 excite waves that enter and exit the system via two distinct ports without significant interaction with the scatterer or PEC walls. 
\item WS modes \#13, \#16, \#22 and \#36 reflect off a single PEC wall before exiting the system, either via the port through which they entered or a different one. 
\item WS modes \#42, \#56 and \#61 experience the longest time delays due to multiple reflections at the PEC walls or the cylinder. The cylindrical shape of the scatterer also introduces chaotic behaviors; mode \#61 interacts the most with the scatterer and thus experiences the longest time delay. 
\end{enumerate}

It is also observed that modes interacting more with the lossy cylinder are associated with smaller $|\bar{S}_{qq}|$. The $|\bar{S}_{61,61}|$ for mode \#61 is the smallest, indicating that the majority of energy is absorbed by the lossy cylinder.

\section{Conclusion}
\label{sec:conclusion}

WS techniques described in \cite{Patel2020WS1} were generalized to account for material dispersion and losses through the addition of new terms in the energy-overlap integrals for the elements of $\matr{Q}$. The  simultaneous diagonalization of $\matr{Q}$ and $\matr{S}$ still produces frequency stable WS modes that experience well-defined group delays, unaffected by system losses. Analytical and numerical results validated the generalized WS relationship and showed that WS modes in dispersive and lossy systems continue to effectively naturally untangle dispersive, resonant, and ballistic scattering phenomena characterized by different dwell times.

\appendices

\section{Solving the Frequency Derivative of Fields}
\label{Appdix:freq_deri2}

For lossy systems, computation of $\matr{Q}$ via Eq.~\eqref{eq:Q_v_loss} requires the frequency derivative of $\fE_p$ and $\fH_p$. Below it is shown that they can be obtained efficiently without the cost of solving for another linear system.

The electric field $\fE$, due to some excitation $\fE_{inc}$ at the ports, satisfies the vector Helmholtz equation
\begin{align}
- \nabla \times \mu_r^{-1} \nabla \times \fE + \omega^2 \mu_0 \varepsilon_0 \varepsilon_r \fE = 0 \label{eq:app_vHelm}
\end{align}
inside the waveguide domain $\Omega$, $\hat{n} \times \fE = 0$ on PEC walls, and the port boundary condition
\begin{align}
    \nx \nabla \times \fE + \mathcal{P}(\fE) = \mathbf{U}_{inc} \label{eq:app_portBC2}
\end{align}
on port surfaces $\partial \Omega$, where \cite{Liu2002Fast,Jin2015finite}
\begin{align}
&\mathcal{P}(\fE) = - \sum_m \sum_n \gamma_{mn} \mathbf{e}_{mn}^\text{TE} \int_{\partial \Omega} \mathbf{e}_{mn}^\text{TE} \cdot \fE dS \\
&\mathbf{U}_{inc} = \nx \nabla \times \fE_{inc} - \sum_m \sum_n \gamma_{mn} \mathbf{e}_{mn}^\text{TE} \int_{\partial \Omega} \mathbf{e}_{mn}^\text{TE} \cdot \fE_{inc} dS \,.
\end{align}
The TE mode excitation is assumed for simplicity of notation, while the above expressions apply to the general case with both TE and TM excitations as well. Using a standard FEM method with proper basis function $\mathbf{\Phi}$, the discretized linear system reads
\begin{align}
\left( - \mathbf{K} + \omega^2 \mu_0 \varepsilon_0 \mathbf{M} + \mathbf{P} \right) \mathbf{x} = \mathbf{v} \label{eq:app_global_linear}
\end{align}
where
\begin{align}
\left[ \mathbf{K} \right]_{ij} &= \langle \nabla \times \mathbf{\Phi}_i, \mu_r^{-1} \nabla \times \mathbf{\Phi}_j \rangle_{\Omega} \\
\left[ \mathbf{M} \right]_{ij} &= \langle \mathbf{\Phi}_i, \varepsilon_r \mathbf{\Phi}_j \rangle_{\Omega} \\
\left[ \mathbf{P} \right]_{ij} &= - \sum_m \sum_n \langle \mathbf{e}_{mn}^\text{TE} \cdot \mathbf{\Phi}_i, \gamma_{mn}, \mathbf{e}_{mn}^\text{TE} \cdot \mathbf{\Phi}_j \rangle_{\partial \Omega} \\
\left[ \mathbf{v} \right]_{i} &= \langle \mathbf{\Phi}_i, \mathbf{U}_{inc} \rangle_{\partial \Omega} \,,
\end{align}
and $\langle \bm{a}, \bm{b} \rangle_\Omega = \int_\Omega \bm{a} \cdot \bm{b} d\Omega$ denotes inner product.

To solve for $\mathbf{x}'$, taking frequency derivative of Eq.~\eqref{eq:app_global_linear} produces
\begin{align}
\left( - \mathbf{K} + \omega^2 \mu_0 \varepsilon_0 \mathbf{M} + \mathbf{P} \right) \mathbf{x}' &= \mathbf{v}' - \mathbf{P}' \mathbf{x} - 2 \omega \mu_0 \varepsilon_0 \mathbf{M} \mathbf{x} \nonumber \\
&\quad + \mathbf{K}' \mathbf{x} - \omega^2 \mu_0 \varepsilon_0 \mathbf{M}' \mathbf{x} \,. \label{eq:app_dw_global_linear}
\end{align}
Note that $\mathbf{x}$ is known upon solving Eq.~\eqref{eq:app_global_linear}; the RHS of Eq.~\eqref{eq:app_dw_global_linear} can be efficiently assembled based on the fact that $\mathbf{P}'$ needs not be explicitly computed and that $\mathbf{M}, \mathbf{K}', \mathbf{M}'$ are sparse; $\mathbf{K}', \mathbf{M}'$ vanish for nondispersive media; the LHS matrix of Eq.~\eqref{eq:app_dw_global_linear} is identical to that of Eq.~\eqref{eq:app_global_linear}, therefore the matrix factorization can be reused. With knowledge of $\mathbf{x}$, only marginal extra cost is needed to solve Eq.~\eqref{eq:app_dw_global_linear} for $\mathbf{x}'.$

The above approach is much more efficient compared to the straightforward finite-difference scheme, $\mathbf{x}' = (\mathbf{x}(\omega + \delta \omega) - \mathbf{x}(\omega) )/\delta \omega$, which requires double the cost of solving Eq.~\eqref{eq:app_global_linear}.

\ifCLASSOPTIONcaptionsoff
  \newpage
\fi

\bibliographystyle{IEEEtran}
\bibliography{IEEEabrv,ref}

\renewenvironment{IEEEbiography}[1]
  {\IEEEbiographynophoto{#1}}
  {\endIEEEbiographynophoto}

\end{document}